\documentclass[aps,prd,twoside,twocolumn,nofootinbib,showpacs,floatfix]{revtex4-1}
\usepackage{amssymb}
\usepackage{amsmath,amssymb}
\usepackage{graphicx,bm}
\usepackage{slashed}
\usepackage{times}
\usepackage{epstopdf}
\usepackage{ulem} 
\usepackage[usenames]{color}
\usepackage{float}
\usepackage{subfigure}
\usepackage{subfigure}
\usepackage{rotating}
\usepackage{color}
\usepackage{multirow}
\usepackage{dcolumn}
\usepackage{overpic}
\usepackage{booktabs}
\usepackage{makecell}
\usepackage{diagbox}

\renewcommand\sout{\bgroup \color{red} \ULdepth=-.5ex \ULset}

\usepackage[colorlinks, citecolor=blue,anchorcolor=red,menucolor=red,linkcolor=blue,filecolor=red,runcolor=red,urlcolor=blue,frenchlinks=red]{hyperref}
\begin{document}
\title{Spectrum of the doubly charmed molecular pentaquarks in chiral effective field theory}
\author{Kan Chen$^{1,2}$}\email{chenk$_$10@pku.edu.cn}
\author{Bo Wang$^{1,2}$}\email{bo-wang@pku.edu.cn}
\author{Shi-Lin Zhu$^{1,2}$}\email{zhusl@pku.edu.cn}
\affiliation{
$^1$Center of High Energy Physics, Peking University, Beijing 100871, China\\
$^2$School of Physics and State Key Laboratory of Nuclear Physics
and Technology, Peking University, Beijing 100871, China}
\begin{abstract}
We perform a systematic study on the interactions of the
$\Sigma_c^{(*)}D^{(*)}$ systems within the framework of chiral
effective field theory. We introduce the contact term,
one-pion-exchange and two-pion-exchange contributions to describe
the short-, long-, and intermediate-range interactions. The low
energy constants of the $\Sigma_c^{(*)}D^{(*)}$ systems are
estimated from the $N\bar{N}$ scattering data by introducing a quark
level Lagrangian. With three solutions of LECs, all the
$\Sigma_c^{(*)}D^{(*)}$ systems with isospin $I=1/2$ can form bound
states, in which different inputs of LECs may lead to
distinguishable mass spectra. In addition, we also investigate the
interactions of the charmed-bottom $\Sigma_c^{(*)}\bar{B}^{(*)}$,
$\Sigma_b^{(*)}D^{(*)}$, and $\Sigma_b^{(*)}\bar{B}^{(*)}$ systems.
Among the obtained bound states, the bindings become deeper when the
reduced masses of the corresponding systems are heavier.
\end{abstract}
\maketitle

\section{Introduction}\label{sec1}
The existence of the $qqqq\bar{q}$ pentaquarks are proposed by
Gell-Mann and Zweig \cite{GellMann:1964nj,Zweig:1981pd,Zweig:1964jf}
at the birth of the quark model in 1964. In 2003, the LEPS group
\cite{Nakano:2003qx} reported a narrow resonance signal at 1540 MeV
with $S=+1$, called $\Theta^+(1540)$, whose quark component should
be $uudd\bar{s}$. Although further experiments did not confirm this
state \cite{MartinezTorres:2010zzb}, it triggered extensive
theoretical and experimental studies on possible pentaquark states
\cite{Zhu:2004xa,Liu:2014yva}.

In 2015, the LHCb Collaboration \cite{Aaij:2015tga,Aaij:2016phn}
measured the $\Lambda_b^0\rightarrow J/\psi K^-p$ decay process and
reported two hidden-charm pentaquark-like states $P_c(4380)$ and
$P_c(4450)$ in the $J/\psi p$ channel, indicating that these two
states have a minimal quark content of $uudc\bar{c}$. In 2019, the
LHCb Collaboration announced \cite{Aaij:2019vzc} the observation of
three narrow peaks in the $J/\psi p$ invariant mass spectrum. They
found that the $P_c(4450)$ is actually composed of two
substructures, the $P_c(4440)$ and $P_c(4457)$ with 5.4$\sigma$
significance. Moreover, they also reported a new state below the
$\Sigma_c\bar{D}$ threshold, namely the $P_c(4312)$ with 7.3$\sigma$
significance. Before the discovery of the LHCb Collaboration in
2015, several groups had predicted
\cite{Wu:2010jy,Yang:2011wz,Wang:2011rga} the existence of molecular
pentaquarks.

The LHCb experiments keep giving us surprise. Very recently, they
reported the first evidence of a charmonium pentaquark candidate
with strangeness in the $\Xi_b^-\rightarrow J/\psi \Lambda K^-$
decay process~\cite{Aaij:2020gdg}. Its mass and width are determined
to be $4458.8\pm2.9^{+4.7}_{-1.1}$ MeV and $17.3\pm
6.5_{-5.7}^{+8.0}$ MeV, respectively. However, its significance just
exceeds 3$\sigma$ after considering all systematic uncertainties.
Further studies on the $P_{cs}^0$ pentaquark are still needed. As
the strange partner of the $P_c$ pentaquark states, it has been
predicted in Refs.
\cite{Wu:2010jy,Chen:2016ryt,Santopinto:2016pkp,Shen:2019evi,Xiao:2019gjd,Wang:2019nvm,Chen:2015sxa}.
Especially, the mass predicted from chiral effective field theory
agrees very well with the experimental data~\cite{Wang:2019nvm}.

Besides the pentaquarks with hidden-charm quark components, the
existence of the double-charm pentaquarks is also an interesting
topic (see Refs.~\cite{Chen:2016qju,Guo:2017jvc,Liu:2019zoy,Lebed:2016hpi,Esposito:2016noz,Brambilla:2019esw,Hosaka:2016pey}
for reviews of the exotic hadrons). For the double-charm pentaquarks, two straightforward
configurations are the compact $ccqq\bar{q}$ pentaquarks and
$(cqq)$-$(c\bar{q})$ baryon-meson molecular states. Based on the
compact pentaquark configuration, the mass spectra of the
pentaquarks with $QQqq\bar{q}$ ($Q=b$, $c$, and $q=u$, $d$, $s$)
quark components were estimated systematically in the framework of
the color-magnetic interaction model~\cite{Zhou:2018bkn}. The
authors of Ref.~\cite{Park:2018oib} used similar approach to
estimate possible stable pentaquark states. In addition, the chiral
quark model \cite{Yang:2020twg} and QCD sum rule~\cite{Wang:2018lhz}
were exploited to analyze the doubly charmed pentaquark states. For
the case of the latter configuration, some theoretical calculations
were performed in the meson exchange models
\cite{Xu:2010fc,Chen:2017vai,Shimizu:2017xrg}. We can qualitatively
capture some features of the double-charm pentaquarks from above
works, while a systematic study of the $\Sigma_c^{(*)}D^{(*)}$
systems is still absent.

The chiral effective field theory has achieved great success in
describing the interactions of the $NN$ systems
\cite{Bernard:1995dp,Epelbaum:2008ga,Machleidt:2011zz,Meissner:2015wva,Hammer:2019poc,Machleidt:2020vzm}.
It is also a very useful tool to study the interactions of the
two-body hadron systems with heavy flavors
\cite{Liu:2012vd,Xu:2017tsr,Wang:2018atz,Meng:2019ilv,Meng:2019nzy,Wang:2019ato,Wang:2019nvm,Wang:2020dhf,Wang:2020dko,Wang:2020htx}.
In the framework of heavy hadron chiral effective theory, we
consider the one-pion-exchange, two-pion-exchange, and contact
contributions to account for the long-, intermediate-, and
short-range interactions of the $\Sigma_c^{(*)}D^{(*)}$ systems,
respectively. Among them, the one-pion-exchange diagrams can be
easily calculated with the standard procedure. For the
two-pion-exchange box diagrams, Weinberg
\cite{Weinberg:1990rz,Weinberg:1991um} suggested that we should only
consider the contributions from two-particle-irreducible (2PI)
graphs, since the two-particle-reducible (2PR) part can be recovered
by inserting the one-pion-exchange potentials into the
nonperturbative iterative equations. This treatment can be done with
the help of the principle-value integral method. For the low energy
constants (LECs) associated with the contact terms, generally, they
should be fixed from the experimental scattering data or lattice QCD
simulations. In Refs. \cite{Wang:2020dhf,Wang:2019nvm,Meng:2019nzy},
we proposed an approach which can relate the contact effective
potentials derived at the hadron level to those derived at the quark
level, so that the LECs can be determined from the quark model. For
example, to estimate the contributions from the contact terms in the
$\Sigma_c^{(*)}D^{(*)}$ systems, we can derive the contact effective
potentials of the $\Sigma_c^{(*)}D^{(*)}$ systems at the quark
level, the coupling constants in the contact terms can be determined
from the $N\bar{N}$ scattering data. Thus, the contributions from
the unknown contact terms can also be estimated. We have a complete
framework to study the interactions of the $\Sigma_c^{(*)}D^{(*)}$
systems, and which is also used to investigate the interactions of
the charmed-bottom $\Sigma_c^{(*)}\bar{B}^{(*)}$,
$\Sigma_b^{(*)}D^{(*)}$, and $\Sigma_b^{(*)}\bar{B}^{(*)}$ systems.

This paper is organized as follows. In Sec. \ref{sec2}, we present
the effective chiral Lagrangians and the effective potentials. In
Sec. \ref{sec3}, we present our numerical results and discussions.
In Sec. \ref{sec4}, we conclude this work with a short summary. Some
supplemental materials for loop diagrams and the results for
charmed-bottom systems are given in the Appendices \ref{app1} and
\ref{app2}, respectively.

\section{Effective chiral Lagrangians and analytical effective potentials}\label{sec2}

We consider the leading order contact and one-pion-exchange
interactions, and the next-to-leading order two-pion-exchange
contributions to describe the scattering amplitudes of the
$\Sigma_cD$, $\Sigma_c^{*}D$, $\Sigma_c D^*$, and $\Sigma_c^*D^*$
systems. We first briefly introduce the effective Lagrangians for
the pionic and contact interactions.

\subsection{Effective chiral Lagrangians}

In the heavy baryon reduction formalism \cite{Scherer:2002tk}, the
leading order nonrelativistic chiral Lagrangians describing the
interactions between the charmed baryons and pion can be constructed
as
\begin{eqnarray}
\label{LagrangianB}
\mathcal{L}_{B\phi}&=&\text{Tr}\left[\mathcal{\bar{B}}_3\left(iv\cdot
D-\delta_c\right)\mathcal{B}_3\right]+2g_5\text{Tr}\left(\bar{\mathcal{B}}_{3^*}^\mu\mathcal{S}\cdot
u\mathcal{B}_{3^*\mu}\right)\nonumber\\&&-\text{Tr}\left[\bar{\mathcal{B}}^{\mu}_{3^*}\left(i
v\cdot
D-\delta_d\right)\mathcal{B}_{3^*\mu}\right]+2g_1\text{Tr}\left(\bar{\mathcal{B}}_3\mathcal{S}\cdot
u \mathcal{B}_3\right)
\nonumber\\&&+2g_2\text{Tr}\left(\bar{\mathcal{B}}_3\mathcal{S}\cdot u\mathcal{B}_1+\text{H.c.}\right)+\frac{1}{2}\text{Tr}\left[\bar{\mathcal{B}}_1\left(i v\cdot D\right)\mathcal{B}_1\right]\nonumber\\
&&+g_3\text{Tr}\left(\mathcal{\bar{B}}_{3^*}^{\mu}u_\mu\mathcal{B}_3+\text{H.c.}\right)+g_4\text{Tr}\left(\bar{\mathcal{B}}_{3^*}^{\mu}u_\mu\mathcal{B}_1+\text{H.c.}\right),\nonumber\\
\end{eqnarray}
where $\mathcal{S}^{\mu}=\frac{i}{2}\gamma_5\sigma^{\mu\nu}v_{\nu}$
is the operator for spin-$\frac{1}{2}$ baryon. The covariant
derivative is defined as $D_\mu\psi=\partial_\mu
\psi+\Gamma_\mu\psi+\psi\Gamma_\mu^T$, where $\Gamma_\mu^T$ is the
transposition of $\Gamma_\mu$. The chiral connection $\Gamma_\mu$
and axial current $u_\mu$ are defined as
\begin{eqnarray}\label{GU}
\Gamma_\mu=\frac{1}{2}\left[\xi^\dagger,\partial_\mu
\xi\right],\quad
u_\mu=\frac{i}{2}\left\{\xi^\dagger,\partial_\mu\xi\right\},
\end{eqnarray}
with
\begin{eqnarray}
\xi^2=U=\text{exp}\left(\frac{i\phi}{f_\pi}\right), \quad
\phi=\left(
                                                                  \begin{array}{cc}
                                                                    \pi^0 & \sqrt{2}\pi+ \\
                                                                    \sqrt{2}\pi^- & -\pi^0 \\
                                                                  \end{array}
                                                                \right).
\end{eqnarray}
Here, $f_\pi=92.4$ MeV is the pion decay constant.

The charmed baryons $\Lambda_c$ and $\Sigma_c^{(*)}$ form the SU(2)
isosinglet and isotriplets, respectively. The spin-$\frac{1}{2}$
isosinglet is
\begin{eqnarray}
\psi_1=\left(
  \begin{array}{cc}
    0 & \Lambda_c^+ \\
    -\Lambda_c^+ & 0 \\
  \end{array}
\right),\quad
\end{eqnarray}
the isotriplet with spin-$\frac{1}{2}$ and spin-$\frac{3}{2}$ are
labeled as $\psi_3$ and $\psi_{3^*}^{\mu}$, respectively. They have
the matrix form
\begin{eqnarray}
\psi_3=\left(
  \begin{array}{cc}
    \Sigma_c^{++} & \frac{\Sigma_c^+}{\sqrt{2}} \\
    \frac{\Sigma_c^+}{\sqrt{2}} & \Sigma_c^0\\
  \end{array}
\right),\quad \psi^{\mu}_{3^*}=\left(
  \begin{array}{cc}
    \Sigma_c^{*++} & \frac{\Sigma_c^{*+}}{\sqrt{2}} \\
    \frac{\Sigma_c^{*+}}{\sqrt{2}} & \Sigma_c^{*0} \\
  \end{array}
\right)^{\mu}.
\end{eqnarray}
The heavy baryon field can be decomposed into the light and heavy
components $\mathcal{B}_i$ and $\mathcal{H}_i$, which read
\begin{eqnarray}
\label{B} \mathcal{B}_i=e^{iM_iv\cdot
x}\frac{1+\slashed{v}}{2}\psi_i,\quad \label{H}
\mathcal{H}_i=e^{iM_iv\cdot x}\frac{1-\slashed{v}}{2}\psi_i,
\end{eqnarray}
where $\psi_i$ denote the heavy baryon fields $\psi_1$, $\psi_3$,
and $\psi_{3^*}$. $v_\mu=\left(1,\textbf{0}\right)$ is the
four-velocity of heavy baryon. The $\mathcal{B}_i$ fields contribute
at the leading order, whereas the $\mathcal{H}_i$ are supressed by
power of $1/m_Q$.
$M_i$ are the masses of the heavy baryons. In this work, we adopt
the following mass splittings \cite{Zyla:2020zbs}
\begin{eqnarray}
\delta_a&=&M_{3^*}-M_3\simeq65\; \rm{MeV},\nonumber\\
\delta_c&=&M_3-M_1\simeq168.5\; \rm{MeV},\nonumber\\
\delta_d&=&M_{3^*}-M_1\simeq233.5\; \rm{MeV}.
\end{eqnarray}

In Eq. \eqref{LagrangianB}, the couplings $g_2=-0.60$ and $g_4=1.04$
can be calculated from the partial decay widths of the
$\Sigma_c\rightarrow \Lambda_c\pi$ and
$\Sigma_c^*\rightarrow\Lambda_c\pi$ processes \cite{Zyla:2020zbs},
respectively. The $g_1$, $g_3$, and $g_5$ can be related to $g_2$
via the quark model \cite{Meguro:2011nr,Liu:2011xc,Meng:2018gan},
which read
\begin{eqnarray}
g_1=0.98, \quad g_3=0.85, \quad g_5=-1.47.
\end{eqnarray}

The leading order chiral Lagrangians for the interactions between
the charmed mesons and pion are \cite{Manohar:2000}
\begin{eqnarray}
\mathcal{L}_{H\phi}&=&-\left\langle\left(i v\cdot
\partial\mathcal{H}\right)\bar{\mathcal{H}}\right\rangle+\left\langle\mathcal{H}
v\cdot \Gamma
\bar{\mathcal{H}}\right\rangle+g\left\langle\mathcal{H}\slashed{u}\gamma_5\bar{\mathcal{H}}\right\rangle\nonumber\\&&-\frac{1}{8}\delta_b\left\langle\mathcal{H}\sigma^{\mu\nu}\bar{\mathcal{H}}\sigma_{\mu\nu}\right\rangle,
\end{eqnarray}
where $\delta_b=m_{D^*}-m_{D}=142.0$ MeV \cite{Zyla:2020zbs}.
$g=-0.59$ represents the axial coupling constant, its value is
calculated from the partial decay width of $D^{*+}\rightarrow
D^0\pi^+$ process \cite{Zyla:2020zbs} and its sign is determined
from the quark model.

In the above Lagrangian, the $\mathcal{H}$ denotes the super-field
of the $\left(D,D^*\right)$ doublet in the heavy quark limit,
\begin{eqnarray}
\mathcal{H}&=&\frac{1+\slashed{v}}{2}\left(P^*_\mu\gamma^\mu+iP\gamma_5\right),\nonumber\\
\bar{\mathcal{H}}&=&\gamma^0H^{\dagger}\gamma^0=\left(P^{*\dagger}_\mu\gamma^\mu+iP^\dagger\gamma_5\right)\frac{1+\slashed{v}}{2},\nonumber\\
P&=&\left(D^0,D^+\right),\qquad
P^*_{\mu}=\left(D^{*0},D^{*+}\right)_\mu.
\end{eqnarray}

Accordingly, the mass splittings for the bottom baryons and mesons
are \cite{Zyla:2020zbs}
\begin{eqnarray}
\delta_a&=&m_{\Sigma_b^*}-m_{\Sigma_b}\simeq 20\;\rm{MeV},\nonumber\\
\delta_b&=&m_{B^*}-m_B\simeq 45\; \rm{MeV},\nonumber\\
\delta_c&=&m_{\Sigma_b}-m_{\Lambda_b}\simeq 191\;\rm{MeV},\nonumber\\
\delta_d&=&m_{\Sigma_b^*}-m_{\Lambda_b}\simeq 211\;\rm{MeV}.
\end{eqnarray}

In the bottom sector, the axial coupling $g=-0.52$ is taken from the
Lattice QCD calculations \cite{Ohki:2008py,Detmold:2012ge}.
$g_2=-0.51$ and $g_4=0.91$ are obtained from the partial decay
widths of the $\Sigma_b\rightarrow \Lambda_b\pi$ and
$\Sigma_b^*\rightarrow \Lambda_b\pi$ \cite{Zyla:2020zbs}. Similarly,
$g_1$, $g_3$, and $g_5$ are determined from the quark model
\cite{Meguro:2011nr,Liu:2011xc,Meng:2018gan},
\begin{eqnarray}
g_1=0.83, \quad g_3=0.72, \quad g_5=-1.25.
\end{eqnarray}

In order to describe the contact interactions of $\Sigma_c^{(*)}$
and $D^{(*)}$, we construct the following Lagrangians,
\begin{eqnarray}
\mathcal{L}_{HB}&=&D_a\left\langle\mathcal{H}\bar{\mathcal{H}}\right\rangle\text{Tr}\left(\bar{\psi}^\mu\psi_\mu\right)\nonumber\\&&+iD_b\epsilon_{\sigma\mu\nu\rho}v^\sigma\left\langle\mathcal{H}\gamma^\rho\gamma_5\bar{\mathcal{H}}\right\rangle
\text{Tr}\left(\bar{\psi}^\mu\psi^\nu\right)\nonumber\\&&+E_a\left\langle\mathcal{H}\tau^i\bar{\mathcal{H}}\right\rangle\text{Tr}\left(\bar{\psi}^\mu\tau_i\psi_{\mu}\right)\nonumber\\&&+iE_b\epsilon_{\sigma\mu\nu\rho}v^\sigma
\left\langle\mathcal{H}\gamma^\rho\gamma_5\tau^i\bar{\mathcal{H}}\right\rangle\text{Tr}\left(\bar{\psi}^\mu\tau_i\psi^\nu\right),\label{Lcontact}
\end{eqnarray}
where
\begin{eqnarray}
\psi^{\mu}&=&\mathcal{B}_{3^*}^\mu-\frac{1}{\sqrt{3}}\left(\gamma^\mu+v^\mu\right)\gamma^5\mathcal{B}_3,\nonumber\\
\bar{\psi}^\mu&=&\bar{\mathcal{B}}_{3^*}^\mu+\frac{1}{3}\bar{\mathcal{B}}_3\gamma^5\left(\gamma^\mu+v^{\mu}\right)
\end{eqnarray}
denote the super-fields of $(\mathcal{B}_{3},\mathcal{B}_{3^*})$
doublet \cite{Cho:1992cf,Cho:1992gg}. The $D_a$, $D_b$, $E_a$, and
$E_b$ are the low energy constants that account for the central
potential, spin-spin interaction, isospin-isospin interaction, and
isospin related spin-spin interaction, respectively. In Sec.
\ref{sec3}, we will use the LECs fitted from the $N\bar{N}$
scattering data \cite{Kang:2013uia} to estimate the contributions of
the leading order contact terms in the $\Sigma_c^{(*)}D^{(*)}$
systems.


\subsection{Effective potentials}\label{Epotential}
To obtain the effective potentials in the momentum space, we first
calculate the scattering amplitude $\mathcal{M}$. The scattering
amplitude $\mathcal{M}$ is related to the effective potential
$\mathcal{V}(\bm{q})$ by the following relation
\begin{eqnarray}
\mathcal{V}\left(\bm{q}\right)=-\frac{\mathcal{M}}{\sqrt{\prod_{i=1}^4
2M_i}},
\end{eqnarray}
where $M_i$ are the masses of the scattering particles. We can
obtain the effective potential $\mathcal{V}(r)$ in the coordinate
space via the following Fourier transformation,
\begin{eqnarray}
\mathcal{V}\left(r\right)&=&\int
\frac{d^3\bm{q}}{\left(2\pi\right)^3}e^{-i\bm{q}\cdot\bm{r}}\mathcal{V}\left(\bm{q}\right)\mathcal{F}\left(\bm{q}\right).
\end{eqnarray}
where a Gaussian regulator $\mathcal{F}\left(\bm{q}\right)={\rm exp}
\left(-\bm{q}^{2n}/\Lambda^{2n}\right)$ is introduced to regularize
the divergence in this integral. This type of regulator has been
widely used in the $NN$ and $N\bar{N}$ systems
\cite{Machleidt:2011zz,Epelbaum:2014efa,Epelbaum:2003xx,Kang:2013uia,Entem:2003ft}.
In this work, we use the LECs fitted from the $N\bar{N}$
scattering~\cite{Kang:2013uia} to estimate the LECs of the
$\Sigma_c^{(*)}D^{(*)}$ systems, thus we use $n=3$ as adopted in
Ref. \cite{Kang:2013uia} for consistency, and take a typical cutoff
$\Lambda=0.4$ GeV to suppress the contributions from higher momenta
\cite{Wang:2020dhf}.

\begin{figure}[htb]
\center
\includegraphics[width=0.7\linewidth]{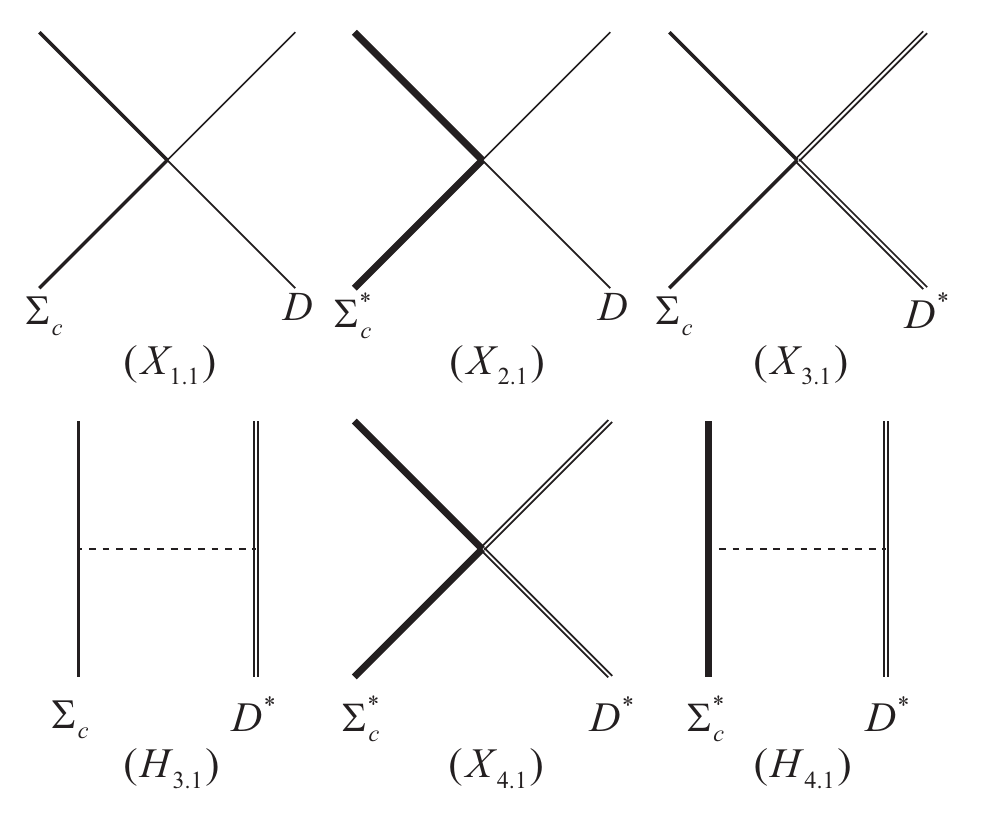}\\
\caption{The leading order Feynman diagrams for the $\Sigma_cD$
($X_{1.1}$), $\Sigma_c^*D$ ($X_{2.1}$), $\Sigma_c D^*$ ($X_{3.1}$,
$H_{3.1}$), and $\Sigma_c^*D^*$ ($X_{4.1}$, $H_{4.1}$) systems. We
use the thick line, heavy-thick line, thin line, double-thin line,
and dashed line to denote the $\Sigma_c$, $\Sigma_c^*$, $D$, $D^*$,
and $\pi$, respectively.}\label{contone}
\end{figure}

The contact and one-pion-exchange interactions contribute to the
leading order effective potentials. The corresponding Feynman
diagrams are collected in Fig. \ref{contone}, where the $\Sigma_c D$
and $\Sigma_c^* D$ systems do not have the one-pion-exchange
diagrams due to the forbidden $DD\pi$ vertex.
The explicit expressions of the contact potentials for the $\Sigma_c
D$, $\Sigma_c^* D$, $\Sigma_c D^*$, and $\Sigma_c^* D^*$ systems are
\begin{eqnarray}
\label{contact1}
\mathcal{V}_{\Sigma_c D}^{X_{1.1}}&=&-D_a+2\left(\mathbf{I}_1\cdot\mathbf{I}_2\right)E_a,\\
\label{contact2}
\mathcal{V}_{\Sigma_c^* D}^{X_{2.1}}&=&-D_a+2\left(\mathbf{I}_1\cdot\mathbf{I}_2\right)E_a,\\
\label{contact3}
\mathcal{V}^{X_{3.1}}_{\Sigma_c D^*}&=&-D_a+2\left(\mathbf{I}_1\cdot\mathbf{I}_2\right)E_a\nonumber\\&&+\frac{2}{3}\left[-D_b+2E_b\left(\mathbf{I}_1\cdot\mathbf{I}_2\right)\right]\bm{\sigma}\cdot \bm{T},\\
\label{contact4} \mathcal{V}^{X_{4.1}}_{\Sigma_c^*
D^*}&=&-D_a+2\left(\mathbf{I}_1\cdot\mathbf{I}_2\right)E_a\nonumber\\&&+\left[-D_b+2E_b\left(\mathbf{I}_1\cdot\mathbf{I}_2\right)\right]\bm{\sigma}_{rs}\cdot
\bm{T}.
\end{eqnarray}
where $\mathbf{I}_1$ and $\mathbf{I}_2$ are the isospin operators of
the $\Sigma_c^{(*)}$ and $D^{(*)}$, respectively. The matrix
elements of $\mathbf{I}_1\cdot\mathbf{I}_2$ can be obtained via
\begin{eqnarray}
\left\langle\mathbf{I}_1\cdot\mathbf{I}_2\right\rangle=\frac{1}{2}\left[I(I+1)-I_1(I_1+1)-I_2(I_2+1)\right],
\end{eqnarray}
where $I$ is the total isospin of the $\Sigma_c^{(*)}D^{(*)}$
systems. The cross product of the final and initial polarization
vectors for $D^*$ mesons ($\boldsymbol{\varepsilon}^\dagger$ and
$\boldsymbol{\varepsilon}$, respectively) is given in terms of
$\bm{T}$ operator
\begin{eqnarray}
-i\bm{T}=\boldsymbol{\varepsilon}^\dagger\times\boldsymbol{\varepsilon}.
\end{eqnarray}
where the spin operator $\bm{S}_v$ of the $D^*$ meson can be related
to the $\bm T$ operator via
\begin{eqnarray}
\bm{S}_v=-\bm{T}. \label{SVT}
\end{eqnarray}
The spin operators $\bm S_s$ of $\Sigma_c$ and $\bm{S}_{rs}$ of
$\Sigma_c^\ast$ are related to the Pauli matrix $\bm{\sigma}$ and
$\bm{\sigma}_{rs}$ via
\begin{eqnarray}
\bm{S}_s=\frac{1}{2}\bm{\sigma},\qquad\bm{S}_{rs}=\frac{3}{2}\bm{\sigma}_{rs}.
\label{SPL}
\end{eqnarray}
Then the matrix elements of the $\bm{\sigma}\cdot\bm{T}$ and
$\bm{\sigma}_{rs}\cdot\bm{T}$ can be obtained from Eqs.
\eqref{SVT}-\eqref{SPL} by
\begin{eqnarray}
\bm{\sigma}\cdot\bm{T}&=&-2\bm{S}_1\cdot\bm{S}_2\nonumber\\&=&-\left[S(S+1)-S_1(S_1+1)-S_2(S_2+1)\right],\nonumber\\
\bm{\sigma}_{rs}\cdot\bm{T}&=&-\frac{2}{3}\bm{S}_{1}\cdot\bm{S}_2\nonumber\\&=&-\frac{1}{3}\left[S(S+1)-S_1(S_1+1)-S_2(S_2+1)\right].\nonumber\\
\end{eqnarray}
where $\bm{S}_1\equiv \bm{S}_s(\bm{S}_{rs})$ and $\bm{S}_2\equiv
\bm{S}_v$ denote the spin operators of the $\Sigma^{(*)}_c$ baryon
and $D^{*}$ meson, respectively.

The one-pion-exchange diagrams for the $\Sigma_c D^*$ and
$\Sigma_c^* D^*$ systems are depicted in graphs ($H_{3.1}$) and
($H_{4.1}$) of Fig. \ref{contone}. The corresponding effective
potentials read
\begin{eqnarray}
\mathcal{V}_{\Sigma_c D^*}^{H_{3.1}}&=&\left(\mathbf{I}_1\cdot\mathbf{I}_2\right)\frac{gg_1}{2f_{\pi}^2}\frac{\left(\bm{q}\cdot\bm{\sigma}\right)\left(\bm{q}\cdot\bm{T}\right)}{\bm{q}^2+m_{\pi}^2},\\
\mathcal{V}_{\Sigma_c^*
D^*}^{H_{4.1}}&=&-\left(\mathbf{I}_1\cdot\mathbf{I}_2\right)\frac{gg_5}{2f_\pi^2}\frac{\left(\bm{q}\cdot\bm{\sigma}_{rs}\right)\left(\bm{q}\cdot\bm{T}\right)}{\bm{q}^2+m_{\pi}^2}.
\end{eqnarray}
One can notice that there is a minus sign between the
one-pion-exchange amplitudes of the $\Sigma_c^{(*)}D^{*}$ and
$\Sigma_c^{(*)}\bar{D}^{*}$ systems~\cite{Wang:2019ato}. This minus
sign comes from the $G$-parity transformation between the
($\bar{D}^{(*)0}$, $D^{(*)-}$) and ($D^{(*)+}$, $D^{(*)0}$)
doublets.
\begin{figure*}[htb]
\center
\includegraphics[width=0.8\linewidth]{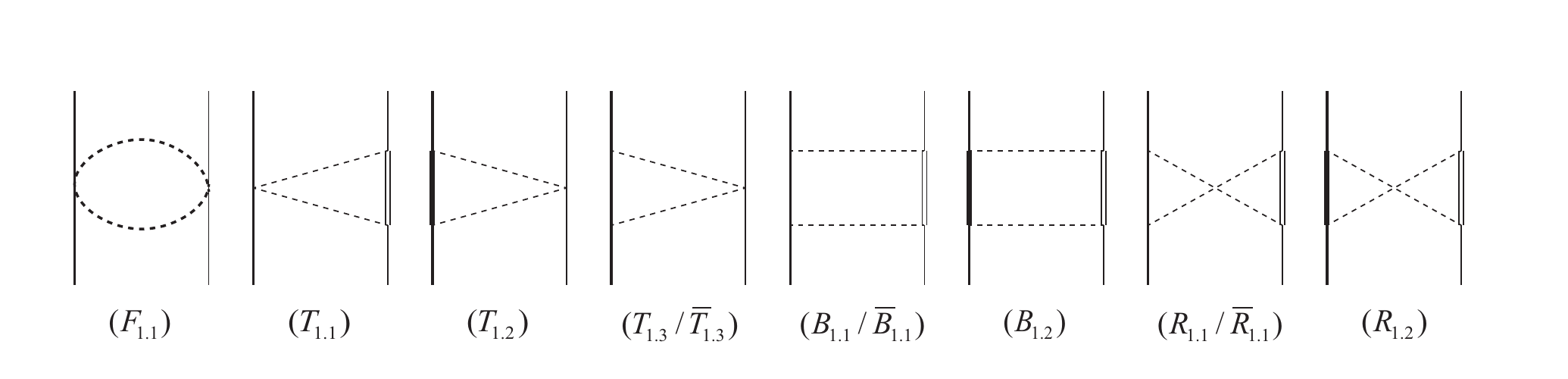}\\
\caption{Two-pion-exchange diagrams that account for the effective
potentials of the $\Sigma_cD$ system at next-to-leading order. These
diagrams include the football diagram ($F_{1.1}$), triangle diagrams
($T_{1.i}/\bar{T}_{1.i}$), box diagrams ($B_{1.i}/\bar{B}_{1.i}$),
and crossed box diagrams ($R_{1.i}/\bar{R}_{1.i}$). The
$\bar{T}_{1.3}$, $\bar{B}_{1.1}$, and $\bar{R}_{1.1}$ denote the
diagrams with $\Lambda_c$ as the intermediate state. The notations
are the same as those in Fig. \ref{contone}.}\label{SigmaDfig}
\end{figure*}
\begin{figure*}[htbp]
\center
\includegraphics[width=0.8\linewidth]{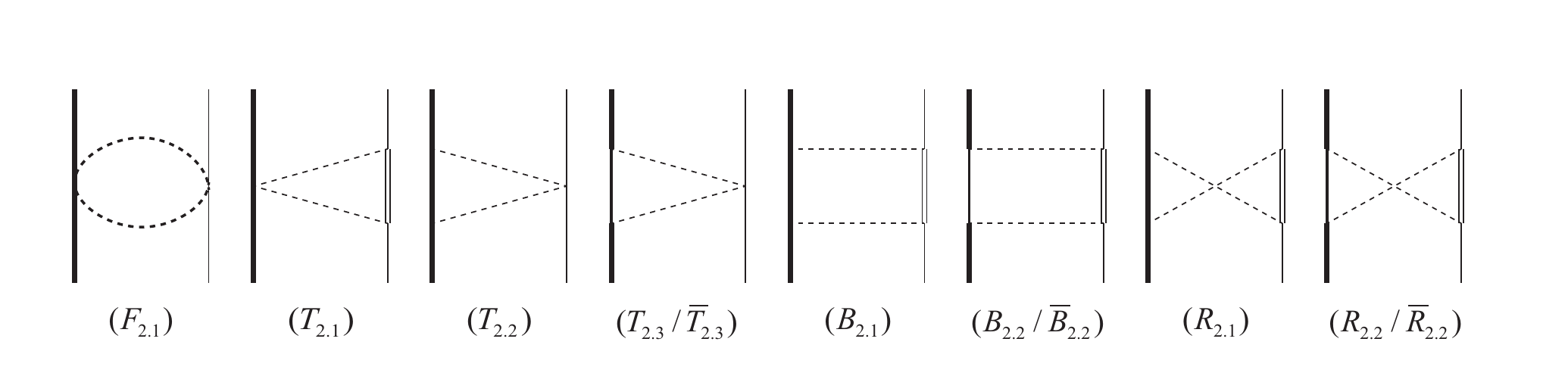}\\
\caption{Two-pion-exchange diagrams for the $\Sigma^*_cD$ system.
The notations are the same as those in Fig.
\ref{SigmaDfig}.}\label{SigmastDfig}
\end{figure*}
\begin{figure*}[htbp]
\includegraphics[width=0.8\linewidth]{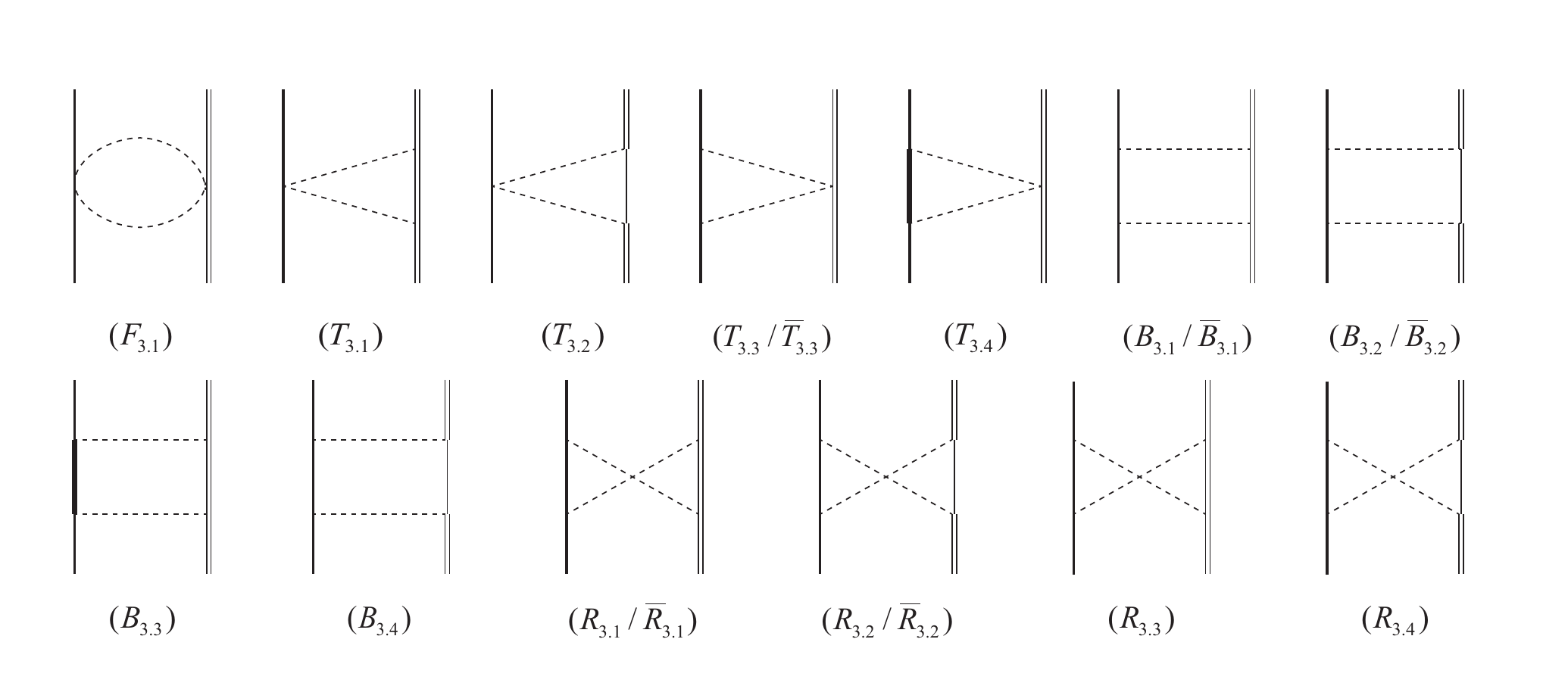}\\
\caption{Two-pion-exchange diagrams for the $\Sigma_cD^*$ system.
The notations are the same as those in Fig.
\ref{SigmaDfig}.}\label{SigmaDstfig}
\end{figure*}
\begin{figure*}[htbp]
\includegraphics[width=0.8\linewidth]{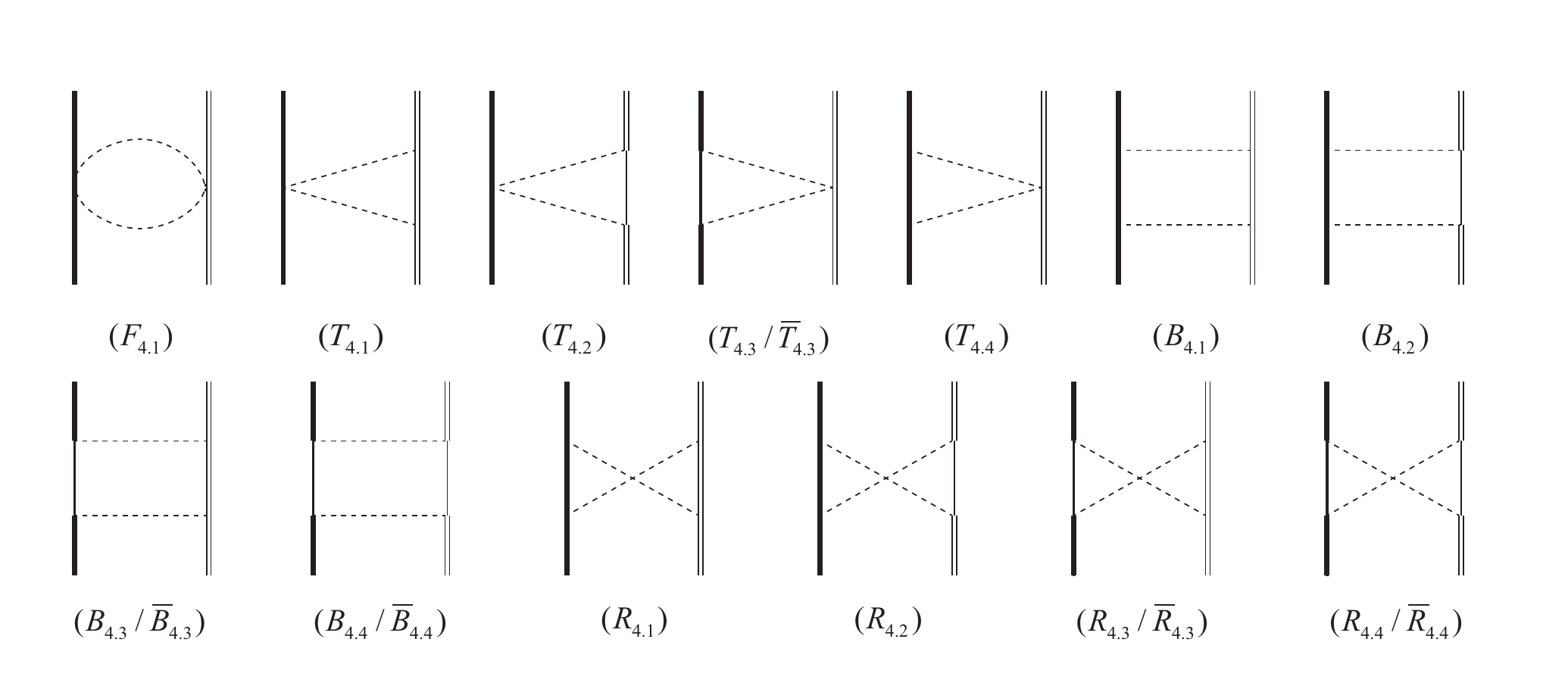}\\
\caption{Two-pion-exchange diagrams for the $\Sigma^*_cD^\ast$
system. The notations are the same as those in Fig.
\ref{SigmaDfig}.}\label{SigmastDstfig}
\end{figure*}

The two-pion-exchange diagrams for the $\Sigma_c D$, $\Sigma_c^*D$,
$\Sigma_c D^*$, and $\Sigma_c^*D^*$ systems are illustrated in Figs.
\ref{SigmaDfig}, \ref{SigmastDfig}, \ref{SigmaDstfig}, and
\ref{SigmastDstfig}, respectively. The analytical results for the
football diagrams ($F_{i.j}$), triangle diagrams ($T_{i.j}$), box
diagrams ($B_{i.j}$), and crossed box diagrams ($R_{i.j}$) generally
have the following forms,

\begin{eqnarray}
\mathcal{V}_{\rm{sys}}^{F_{i.j}}&=&\left(\mathbf{I}_1\cdot\mathbf{I}_2\right)\frac{1}{f_\pi^4}J_{22}^F(m_\pi,q),\label{football}\\
\mathcal{V}_{\rm{sys}}^{T_{i,j}}&=&\left(\mathbf{I}_1\cdot\mathbf{I}_2\right)\frac{\mathcal{C}_{\rm{sys}}^{T_{i.j}}}{f_\pi^4}
\left[-\bm{q}^2\mathcal{C}_1^{T_{i.j}}\left(J_{24}^T+J_{33}^T\right)\right.\nonumber\\&&\left.+\mathcal{C}_2^{T_{i.j}}J_{34}^T\right]\left(m_\pi,\mathcal{E}^{T_{i.j}},q\right),
\label{triangle}\\
\mathcal{V}_{\rm{sys}}^{B_{i.j}}&=&\left(1-\mathbf{I}_1\cdot\mathbf{I}_2\right)\frac{\mathcal{C}_{\rm{sys}}^{B_{i.j}}}{f_\pi^4}\left[-\bm{q}^2\mathcal{C}_1^{B_{i.j}}J_{21}^B\right.\nonumber\\
&&+\bm{q}^4\mathcal{C}_2^{B_{i.j}}(J_{22}^B+2J_{32}^B+J_{43}^B)-\bm{q}^2\mathcal{C}_3^{B_{i.j}}(J_{31}^B\nonumber\\&&+J_{42}^B)\left.+\mathcal{C}_4^{B_{i.j}}J_{41}^B
\right]\left(m_\pi,\mathcal{E}_1^{B_{i.j}},\mathcal{E}^{B_{i.j}}_2,q\right),
\label{box}\\
\mathcal{V}_{\rm{sys}}^{R_{i.j}}&=&\left(1+\mathbf{I}_1\cdot\mathbf{I}_2\right)\frac{\mathcal{C}_{\rm{sys}}^{R_{i.j}}}{f_\pi^4}\left[-\bm{q}^2\mathcal{C}_1^{R_{i.j}}J_{21}^R\right.\nonumber\\&&
+\bm{q}^4\mathcal{C}_2^{R_{i.j}}\left(J_{22}^R+2J_{32}^R+J_{43}^R\right)-\bm{q}^2\mathcal{C}_3^{R_{i.j}}(J_{31}^R\nonumber\\&&+J_{42}^R)
\left.+\mathcal{C}_4^{R_{i.j}}J_{41}^R
\right]\left(m_\pi,\mathcal{E}_1^{R_{i.j}},\mathcal{E}^{R_{i.j}}_2,q\right),
\label{crossedbox}
\end{eqnarray}
where the subscript ``sys" denotes the corresponding
$\Sigma_c^{(*)}D^{(*)}$ system. The superscripts $F_{i.j}$,
$T_{i.j}$, $B_{i.j}$, and $R_{i.j}$ are the labels of Feynman
diagrams illstrated in Figs. \ref{SigmaDfig}-\ref{SigmastDstfig}.
The $J_x^T$, $J_x^B$, $J_x^R$ are scalar loop functions defined in
Appendix A of Ref. \cite{Wang:2019ato}. $\mathcal{E}^{T_{i.j}}$,
$\mathcal{E}_{1(2)}^{B_{i.j}}$, $\mathcal{E}^{R_{i.j}}_{1(2)}$ are
the residual energies.

In Ref. \cite{Wang:2019ato}, we found the contributions of
$\Lambda_c$ in the loops of two-pion-exchange diagrams have
considerable corrections to the effective potentials. Thus, in this
work, we also consider the contributions from intermediate
$\Lambda_c$ state. The general expressions for the corresponding
triangle diagrams ($\bar{T}_{i.j}$), box diagrams ($\bar{B}_{i.j}$),
and crossed box diagrams ($\bar{R}_{i.j}$) read
\begin{eqnarray}
\mathcal{V}^{\bar{T}_{i.j}}_{\rm{sys}}&=&\left(\mathbf{I}_1\cdot\mathbf{I}_2\right)\frac{C_{\rm{sys}}^{\bar{T}_{i.j}}}{f_\pi^4}\left[-\bm{q}^2\mathcal{C}_1^{\bar{T}_{i.j}}\left(J_{24}^T+J_{33}^T\right)\right.\nonumber\\&&\left.
+\mathcal{C}_2^{\bar{T}_{i.j}}\right]\left(m_\pi,\mathcal{E}^{\bar{T}_{i.j}},q\right),
\label{Ltriangle}\\
\mathcal{V}_{\rm{sys}}^{\bar{B}_{i.j}}&=&\left(1-2\mathbf{I}_1\cdot\mathbf{I}_2\right)\frac{\mathcal{C}_{\rm{sys}}^{\bar{B}_{i.j}}}{f_\pi^4}\left[-\bm{q}^2\mathcal{C}_{1}^{\bar{B}_{i.j}}J_{21}^B\right.\nonumber\\&&
+\bm{q}^4\mathcal{C}_2^{\bar{B}_{i.j}}\left(J_{22}^B+2J_{32}^B+J_{43}^B\right)-\bm{q}^2\mathcal{C}_{3}^{\bar{B}_{i.j}}(J^B_{31}\nonumber\\&&+J_{42}^B)\left.+\mathcal{C}_4^{\bar{B}_{i.j}}J_{41}^B\right]
\left(m_\pi,\mathcal{E}_1^{\bar{B}_{i.j}},\mathcal{E}_2^{\bar{B}_{i.j}},q\right),
\label{Lbox}\\
\mathcal{V}_{\rm{sys}}^{\bar{R}_{i.j}}&=&\left(1+2\mathbf{I}_1\cdot\mathbf{I}_2\right)\frac{\mathcal{C}_{\rm{sys}}^{\bar{R}_{i.j}}}{f_\pi^4}\left[-\bm{q}^2\mathcal{C}_{1}^{\bar{R}_{i.j}}J_{21}^R\right.\nonumber\\&&
+\bm{q}^4\mathcal{C}_2^{\bar{R}_{i.j}}\left(J_{22}^R+2J_{32}^R+J_{43}^R\right)-\bm{q}^2\mathcal{C}_{3}^{\bar{R}_{i.j}}(J^R_{31}\nonumber\\&&+J_{42}^R)\left.+\mathcal{C}_4^{\bar{R}_{i.j}}J_{41}^R\right]
\left(m_\pi,\mathcal{E}_1^{\bar{R}_{i.j}},\mathcal{E}_2^{\bar{R}_{i.j}},q\right),\label{Lcrossedbox}
\end{eqnarray}
One can see Appendix \ref{app1} for the explicit values of the
coefficients defined in Eqs. \eqref{triangle}-\eqref{Lcrossedbox}.

We notice the expressions of the two-pion-exchange diagrams for the
$\Sigma_c^{(*)}D^{(*)}$ systems are identical to those of the
$\Sigma_c^{(*)}\bar{D}^{(*)}$ systems \cite{Wang:2019ato}. This
interesting results can be easily understood as follows: the
differences between the two-pion-exchange amplitudes of the
$\Sigma_c^{(*)}D^{(*)}$ and $\Sigma_c^{(*)}\bar{D}^{(*)}$ systems
are completely caused by the pionic coupling of the charmed and
anti-charmed mesons. As mentioned before, the one-pion vertices
[from the $u_\mu$ in Eq.~\eqref{GU}] between the charmed and
anti-charmed mesons have a minus sign difference, but they appear in
pairs in the two-pion-exchange diagrams. Besides, the two-pion
vertices [from the $\Gamma_\mu$ in Eq.~\eqref{GU}] is invariant
under the $G$-parity transformation.

We have subtracted the 2PR contributions of the box diagrams in our
calculations. This can be achieved by the principal-value integral
method proposed in Ref. \cite{Wang:2019ato}, in which a detailed
derivation is presented in  the Appendix B.

\section{Numerical results and discussions}\label{sec3}

To get the numerical results, we need to determine the four LECs
defined in Eq. \eqref{Lcontact}. At present, there are no
experimental data or lattice QCD simulations for the possible
$P_{cc}$ states. In Refs.
\cite{Wang:2020dhf,Wang:2019nvm,Meng:2019nzy}, we proposed to bridge
the LECs determined from the $NN$ ($N\bar{N}$) scattering data to
the unknown LECs of the di-hadron systems via a quark level contact
Lagrangian. In this work, we apply this approach to estimate the
contributions of the contact terms for the $\Sigma_c^{(*)}D^{(*)}$
systems, likewise. Then we search for binding solutions via solving
the Schr\"odinger equation and discuss the numerical results.
\subsection{Determining the LECs of the $\Sigma_c^{(*)}D^{(*)}$ systems}

It is assumed that the contact terms are mimicked by exchanging
heavy mesons through the $S$-wave
interaction~\cite{Wang:2020dhf,Wang:2019nvm,Meng:2019nzy}, in which
a general quark-level Lagrangian is constructed as
\begin{eqnarray}
\mathcal{L}=g_s\bar{q}\mathcal{S}q+g_a\bar{q}\gamma_\mu\gamma^5\mathcal{A}^\mu
q,\label{Lquark} \label{quarkLL}
\end{eqnarray}
where $q=(u,d)$, $c_s$ and $c_t$ are two independent coupling
constants. The fictitious scalar ($\mathcal{S}$) and axial-vector
($\mathcal{A}^\mu$) fields with positive parity are introduced to
account for the central potential and spin-spin interaction,
respectively. From Eq. \eqref{quarkLL}, the $q\bar{q}$ contact
potential is obtained as
\begin{eqnarray}
V_{q\bar{q}}=c_s\left(1-3\bm{\tau}_1\cdot\bm{\tau}_2\right)+c_t\left(1-3\bm{\tau}_1\cdot\bm{\tau}_2\right)
\bm{\sigma}_1\cdot\bm{\sigma}_2. \label{qqbar}
\end{eqnarray}

In Table \ref{operators}, we present the quark-level matrix elements
of the operators related to the contact potentials.
\begin{table}[!htbp]
\renewcommand\arraystretch{1.7}
\caption{The quark-level matrix elements of two-body interaction
operators $\mathcal{O}_{ij}$ for the $N\bar{N}$ and $\Sigma_c D^*$
systems.\label{operators}} \setlength{\tabcolsep}{2.55mm}{
\begin{tabular}{c|cccccccccccccccccc}
\toprule[1pt]
$\mathcal{O}_{ij}$&$\bm{1}_{ij}$&$\bm{\tau}_i\cdot\bm{\tau}_j$&$\bm{\sigma}_i\cdot\bm{\sigma}_j$&$\left(\bm{\tau}_i\cdot\bm{\tau}_j\right)\left(\bm{\sigma}_i\cdot\bm{\sigma}_j\right)$\\
\hline
$\left[N\bar{N}\right]^{I=1}_{J=1}$&9&1&1&$\frac{25}{9}$\\
$\left[N\bar{N}\right]^{I=1}_{J=0}$&9&1&-3&-$\frac{25}{3}$\\
$\left[N\bar{N}\right]^{I=0}_{J=1}$&9&-3&1&-$\frac{25}{3}$\\
$\left[N\bar{N}\right]^{I=0}_{J=0}$&9&-3&-3&25\\
$\left[\Sigma_cD^*\right]_{J=\frac{3}{2}}^{I=\frac{3}{2}}$&2&2&$\frac{4}{3}$&$\frac{4}{3}$\\
\bottomrule[1pt]
\end{tabular}
}
\end{table}
Based on the Lagrangian in Eq. \eqref{quarkLL} and the matrix
elements in Table \ref{operators}, the authors of Ref.
\cite{Wang:2020dhf} derived the contact potential of the $N\bar{N}$
system with quantum numbers $I=1$ and $^{2S+1}L_J=^3S_1$,
\begin{eqnarray}
V_{N\bar{N}}^{{}^3S_1}&=&\left\langle
N\bar{N}\left|V_{q\bar{q}}\right|N\bar{N}\right\rangle=6c_s-\frac{22}{3}c_t.
\end{eqnarray}
One can as well as obtain the contact potentials of the
$[N\bar{N}]_{J=0}^{I=1}$, $[N\bar{N}]_{J=1}^{I=0}$, and
$[N\bar{N}]_{J=0}^{I=0}$ systems, accordingly.

Similarly, the $\Sigma_c D^*$ contact potential can be obtained from
Eq. \eqref{qqbar} and Table \ref{operators} as
\begin{eqnarray}
V_{\Sigma_c
D^*}&=&2c_s-12c_s\mathbf{I}_1\cdot\mathbf{I}_2-\frac{4}{3}c_t\bm{\sigma}\cdot\bm{T}\nonumber\\&&+8c_t\left(\mathbf{I}_1\cdot\mathbf{I}_2\right)\left(\bm{\sigma}\cdot\bm{T}\right).\label{qqbarSigmaD}
\end{eqnarray}
Comparing Eq. \eqref{contact3} with Eq. \eqref{qqbarSigmaD} we get
\begin{eqnarray}
D_a=-2c_s,\quad E_a=-6c_s,\quad D_b=2c_t,\quad E_b=6c_t.\nonumber\\
\end{eqnarray}

In Ref. \cite{Kang:2013uia}, based on the $N\bar{N}$ scattering
data, the LECs for the $I=0$ ($J=0,1$) and $I=1$ ($J=0,1$)
$N\bar{N}$ systems are fitted. With the LECs of these four
$N\bar{N}$ systems, we obtain six sets of solutions for the $c_s$
and $c_t$. Among them, four sets of $c_s$ and $c_t$ are consistent
with each other in sizes and signs:
\begin{itemize}
  \item []
  Set 1: $c_s=-5.84$ GeV$^{-2}$, ~~~~$c_t=2.50$ GeV$^{-2}$;
  \item []
  Set 2: $c_s=-8.10$ GeV$^{-2}$, ~~~~$c_t=0.65$ GeV$^{-2}$;
  \item []
  Set 3: $c_s=-8.25$ GeV$^{-2}$, ~~~~$c_t=0.52$ GeV$^{-2}$;
  \item []
  Set 4: $c_s=-7.71$ GeV$^{-2}$, ~~~~$c_t=0.38$ GeV$^{-2}$.
\end{itemize}
The remaining two sets of solutions either have the different signs or are too large, leading to unstable numerical results in our calculations. 

When checking the above four sets of LECs, we notice that the $c_s$
value in Set 1 is smaller than those of other sets, the input with
$c_s$ in Set 1 will lead to relatively small central potentials. On
the contrary, the value of $c_t$ in Set 1 is larger than those of
other sets, the spin-spin corrections would be important with this
set of LECs. This is the first case we want to discuss, we label
this set of LECs solution as Case 1. The LECs in Set 2 have been
successfully applied to study the interactions of the $D^{(*)}N$
systems \cite{Wang:2020dhf}, the small value of $c_t$ shows that
with this set of solution, the spin-spin interaction serves as the
perturbation to the $D^{(*)}N$ multiplets. The LECs in Sets 3 and 4
are very close to that of Set 2, and they give very similar results.
Thus, we will use the LECs in Set 2 as our Case 2. In addition, we
also use the least square method to fit a best solution from these
four sets of LECs, the solution are obtained as
\begin{eqnarray}
c_s=-7.46 \rm{GeV}^{-2},\qquad c_t=1.02 \rm{GeV}^{-2}.\label{LECs}
\end{eqnarray}
we label this set of LECs as the Case 3.

\subsection{Numerical results of the effective potentials}
\begin{figure*}[htbp]
\includegraphics[width=0.6\linewidth]{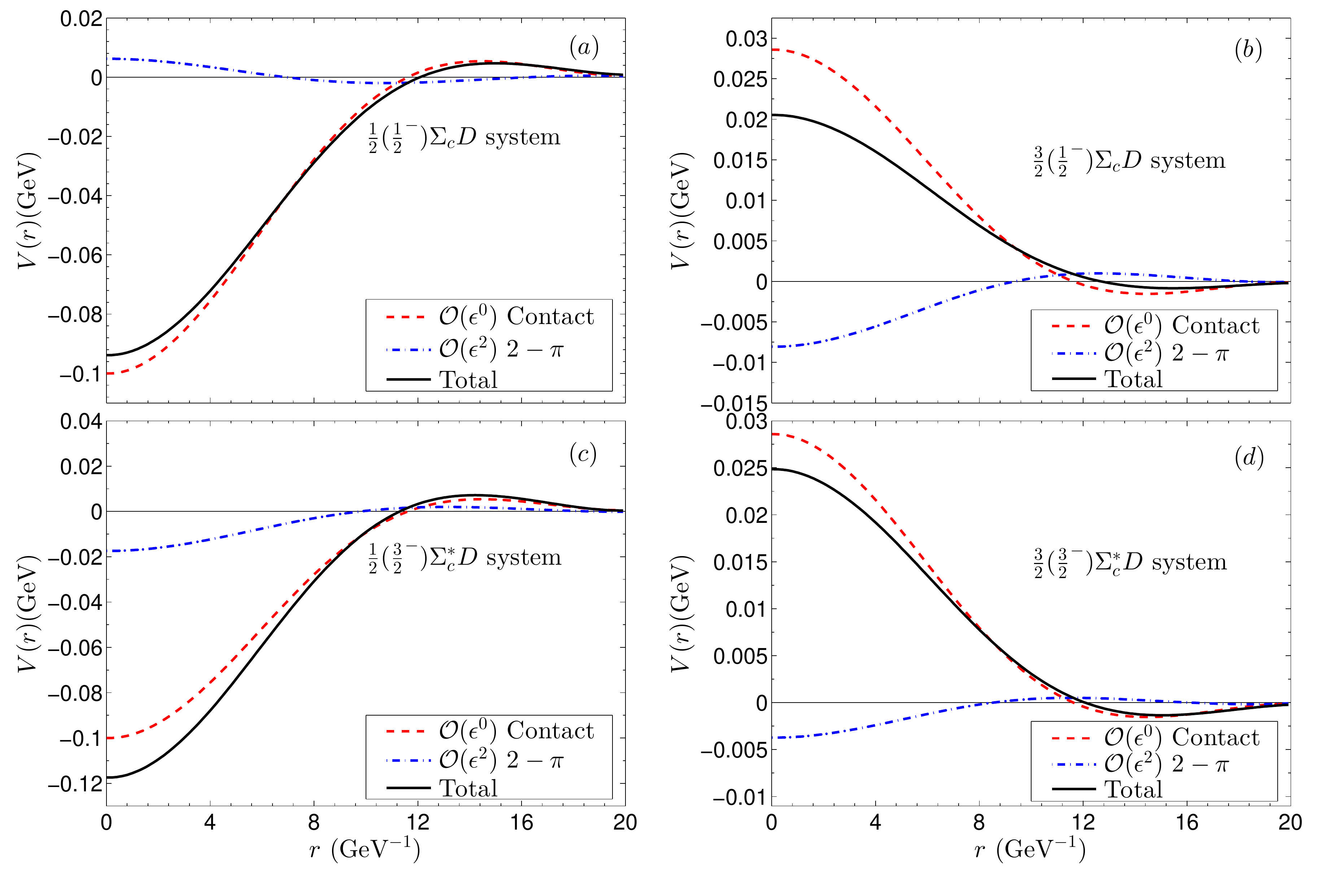}\\
\caption{The effective potentials for the $[\Sigma^{(*)}_c
D]^{I=1/2}_J$ systems. Their $I(J^P)$ numbers are illustrated in
each subfigure. The red dashed line and blue dot-dashed line denote
the effective potentials from the contact term and
two-pion-exchange, respectively. The black solid line denote the
total effective potential for each system. }\label{SgmSgmstDfig}
\end{figure*}
We use the LECs in Case 3 to present the effective potentials of all the $\Sigma_c^{(*)}D^{(*)}$ systems. In Fig. \ref{SgmSgmstDfig}, we plot the effective potentials of the $\Sigma_cD$ and $\Sigma_c^*D$ systems. 
The contact terms of the $\Sigma_cD$ and $\Sigma_c^*D$ are the same
in the heavy quark limit, which can be checked from the line shapes
of the contact effective potentials in Fig. \ref{SgmSgmstDfig}.

For the $[\Sigma_c D]^{I=1/2}_{J=1/2}$ system, the two-pion-exchange
interaction provides a weakly repulsive force. The contact
interaction provides a strong attractive force, which is also true
for the $[\Sigma^{*}_c D]^{I=1/2}_{J=3/2}$ system. In the
$[\Sigma^{*}_c D]^{I=1/2}_{J=3/2}$ system, the two-pion-exchange
interaction provides a weakly attractive potential and forms a
deeply bound $[\Sigma^{*}_c D]^{I=1/2}_{J=3/2}$ state together with
the strong attractive contact term.

From the right panel of Fig. \ref{SgmSgmstDfig}, we can see that the
two-pion-exchange interactions provide considerable attractive force
in the $[\Sigma^{(*)}_c D]^{I=3/2}_J$ systems. However, for the
$I=3/2$ case, the contact terms provide strong repulsive forces and
the total effective potentials are repulsive, i.e., we can not find
any bound states. This is also true for the $[\Sigma^{(*)}_c
D^*]^{I=3/2}_J$ systems. Thus, in the following, we only discuss the
$\Sigma_c^{(*)}D^{(*)}$ systems with $I=1/2$.

\begin{figure*}[htbp]
\includegraphics[width=0.8\linewidth]{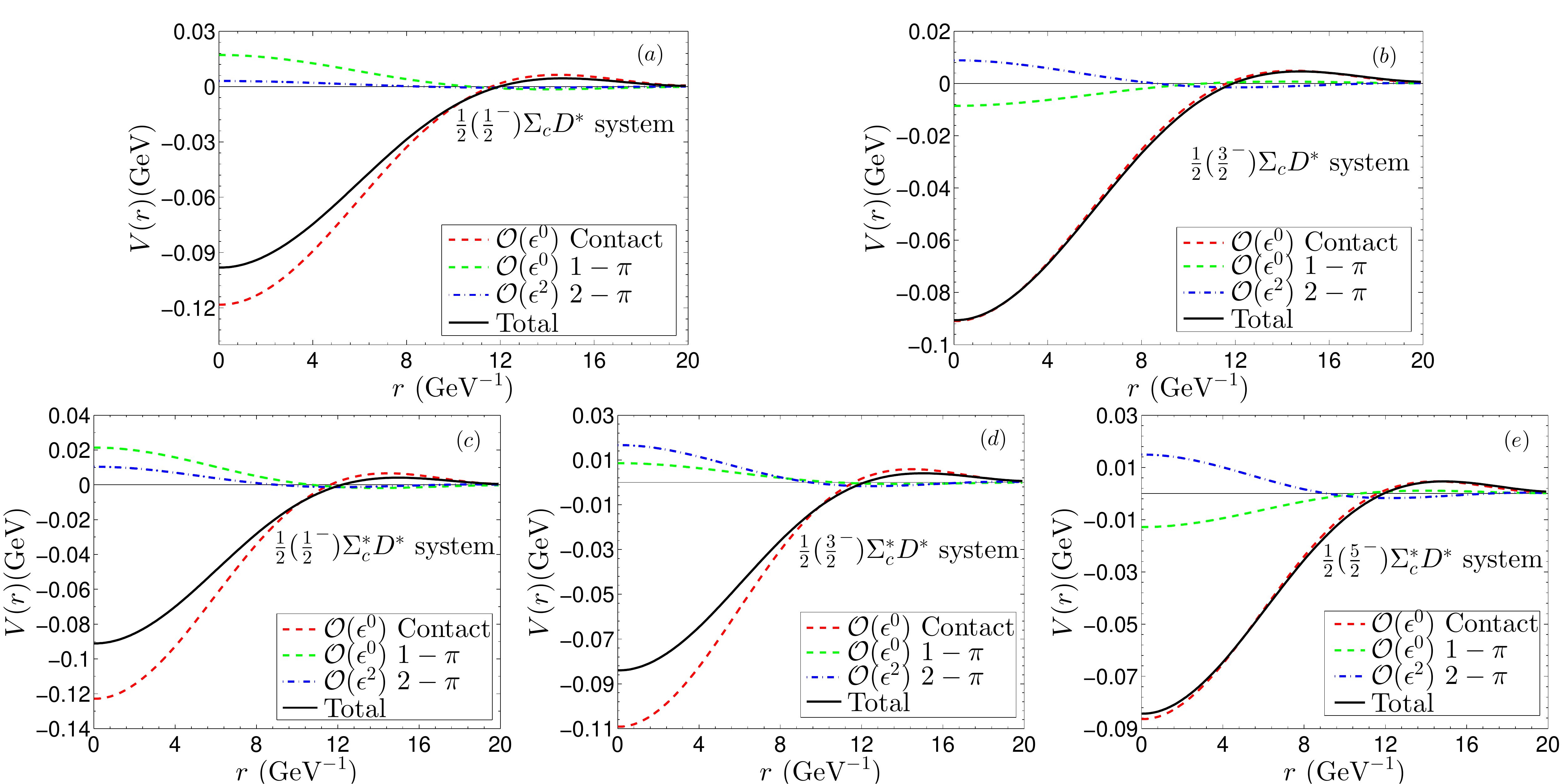}\\
\caption{The effective potentials for the $[\Sigma^{(*)}_c
D^{*}]^{I=1/2}_J$ systems. Their $I(J^P)$ numbers are illustrated in
each subfigure. The red dashed line, green dashed line and blue
dot-dashed line denote the effective potentials from the contact
term, one-pion- and two-pion-exchange interactions, respectively.
The black solid line denote the total effective potential for each
system.}\label{SgmSgmstDstfig}
\end{figure*}

In Fig. \ref{SgmSgmstDstfig}, we present the effective potentials of
the $\Sigma_c^{(*)}D^*$ systems. From Fig. \ref{SgmSgmstDstfig}, we
can see that in the $[\Sigma_cD^*]_J^{I=1/2}$ and
$[\Sigma_c^*D^*]_J^{I=1/2}$ systems, the contact terms provide
strong attractive force. The one-pion- and two-pion-exchange
interactions supply very weak repulsive forces in the
$[\Sigma_cD^*]_{J=1/2}^{I=1/2}$,
$[\Sigma^*_cD^*]_{J=1/2(3/2)}^{I=1/2}$ systems. For the
$[\Sigma_cD^*]_{J=3/2}^{I=1/2}$ and
$[\Sigma_c^*D^*]_{J=5/2}^{I=1/2}$ systems, the one-pion- and
two-pion-exchange potentials supply comparable attractive and
repulsive forces, respectively. Thus, the total effective potentials
for these two systems are nearly equivalent to their contact
potentials.


Our calculation shows that the contact terms are important to the
$[\Sigma_c^{(*)}D^{(*)}]^{I=1/2}_J$ systems. In our framework, the
fictitious scalar mesons account for the mainly attractive
interactions in the $[\Sigma_c^{(*)}D^{(*)}]^{I=1/2}_J$ systems, and
the exchange of axial-vector mesons result in mass splittings in
spin multiplets. In Eq. \eqref{Lquark}, the $c_s$ and $c_t$ are
related to the strength of the scalar-exchange and
axial-vector-exchange forces, respectively. From Table \ref{CC}, we
notice that the axial-vector-exchange interactions which are related
to the $c_t$ provide small corrections to the final effective
potentials in Cases 2 and 3. However, in Case 1, we have
$|c_t/c_s|=0.43$, this can be regarded as our upper limit of the
$|c_t/c_s|$. The relatively small value of $c_t$ can be traced to
the large masses of the axial-vector particles since their masses
exceed $1$ GeV~\cite{Zyla:2020zbs}.

\subsection{The binding energies of the $\Sigma_c^{(*)}D^{(*)}$ systems}\label{BE}

\begin{table*}[!htbp]
\caption{The binding energies, masses and root-mean-square radii for
all the $[\Sigma_c^{(*)}D^{(*)}]^{I=1/2}_J$ systems. The subscript
denotes the total angular momentum of this system. The adopted LECs
in Cases 1, 2, and 3 are ($c_s=-5.84$, $c_t=2.50$) GeV$^{-2}$,
($c_s=-8.10$, $c_t=0.65$ ) GeV$^{-2}$, and ($c_s=-7.46$, $c_t=1.02$)
GeV$^{-2}$, respectively.}\label{CC}
\renewcommand\arraystretch{1.15}
\setlength{\tabcolsep}{4.1mm}{
\begin{tabular}{c|c|ccccccccccccccccc}
\toprule[1pt] &&$\left[\Sigma_cD\right]_{\frac{1}{2}}$
&$\left[\Sigma^*_cD\right]_{\frac{3}{2}}$&$\left[\Sigma_cD^*\right]_{\frac{1}{2}}$&$\left[\Sigma_cD^*\right]_{\frac{3}{2}}$& $\left[\Sigma_c^*D^*\right]_{\frac{1}{2}}$&$\left[\Sigma_c^*D^*\right]_{\frac{3}{2}}$&$\left[\Sigma_c^*D^*\right]_{\frac{5}{2}}$\\
\hline
\multirow{2}{*}{Case 1} &BE (MeV)&-15.4&-25.0&-31.8&-8.0&-32.8&-18.2&-3.5\\
&$R_{rms}$ (fm)&1.45&1.25&1.20&1.65&1.20&1.38&1.91\\
\hline
\multirow{2}{*}{Case 2}&BE (MeV)&-31.3&-42.9&-30.3&-31.7&-26.6&-25.4&-29.7\\
&$R_{rms}$ (fm)&1.23&1.11&1.22&1.20&1.26&1.27&1.22\\
\hline
\multirow{2}{*}{Case 3} &BE (MeV)&-26.5&-37.7&-29.1&-25.0&-26.4&-22.6&-22.2\\
&$R_{rms}$ (fm)&1.27&1.14&1.23&1.27&1.26&1.31&1.30\\
\hline \bottomrule[1pt]
\end{tabular}
}
\end{table*}
The binding energies, masses, and the root-mean-square radii in the
above three cases are presented in Table \ref{CC}. We find bound
state solutions only for the $I=1/2$ channels. The $R_{rms}$ are
about 1$-$2 fm for all the considered $\Sigma_c^{(*)}D^{(*)}$
systems, which are the typical sizes of the hadronic molecules. From
Table \ref{CC}, we can see that the binding of the $[\Sigma_c^*
D]_{J=3/2}^{I=1/2}$ system is deeper than that of the
$[\Sigma_cD]_{J=1/2}^{I=1/2}$ system. In the heavy quark limit, the
$[\Sigma_c D]_{J=1/2}^{I=1/2}$ and $[\Sigma_c^*D]_{J=3/2}^{I=1/2}$
systems share the same contact term. Thus, the difference of the
binding energy is from the contributions of the two-pion-exchange
interactions. Besides, in Case 1, the $\Sigma_c^{(*)}D^*$ systems
with lower total angular momentum $J$ are more compact. This
situation is very similar to the $\Sigma_c^{(*)}\bar{D}^*$ systems
\cite{Wang:2019ato}. However, the results in Cases 2 and 3 show that
the binding energies for the different $\Sigma_c^{(*)}D^{(*)}$
systems are comparable to each other, and they have very similar
spatial sizes.

The parameters $c_s$ and $c_t$ are related to the central potentials
and spin-spin interactions, respectively. In Case 1,
$|c_t/c_s|=0.43$, the spin-spin corrections in contact terms have
considerable contributions in the $\Sigma_c^{(*)}D^{*}$ systems
[note that the spin-spin corrections do not contribute to the
$\Sigma_c^{(*)}D$ systems, e.g., see Eqs.
\eqref{contact1}-\eqref{contact2}]. The spin-spin corrections are
much larger than the contributions from the one-pion-exchange and
two-pion-exchange interactions. Thus, in Case 1, the mass splittings
among different $[\Sigma_c^{(*)}D^{*}]_J^{I=1/2}$ systems are mainly
caused by the corrections of the spin-spin interactions. The results
obtained from the Cases 2 and 3 are close to each other. In contrast
to Case 1, $|c_t/c_s|\approx 0.1$ in these two cases, i.e., the
central potentials are dominant and the spin-spin potentials are
small. The contributions from the spin-spin interactions are
comparable to those of the one-pion-exchange and two-pion-exchange
interactions. As shown in Table \ref{CC}, in these two cases, the
$\Sigma_c^{(*)}D^*$ systems with higher total angular momenta have
deeper binding energies.

The results for the $\Sigma_c^{(*)}\bar{B}^{(*)}$,
$\Sigma_b^{(*)}D^{(*)}$, and $\Sigma_b^{(*)}\bar{B}^{(*)}$ systems
are given in Appendix \ref{app2}.

\subsection{Possible decay patterns of the $\Sigma_c^{(*)}D^{(*)}$ molecules}
The $P_c(4312)$, $P_c(4440)$, and $P_c(4457)$ are produced in the
$\Lambda_b^0\rightarrow J/\psi K^-p$ decay process and reconstructed
in the $J/\psi p$ channel~\cite{Aaij:2019vzc}. Similarly, the
$P_{cs}(4459)$ is produced in the $\Xi_b^-\rightarrow J/\psi K^-
\Lambda$ process and observed in the $J/\psi \Lambda$ invariant mass
spectrum \cite{Aaij:2020gdg}. In this subsection, we discuss the
possible decay patterns of the $P_{cc}$ states. They may be
considered as the reconstructive channels from the $pp$ collisions
at LHCb.

\begin{figure}[htbp]
\includegraphics[width=0.85\linewidth]{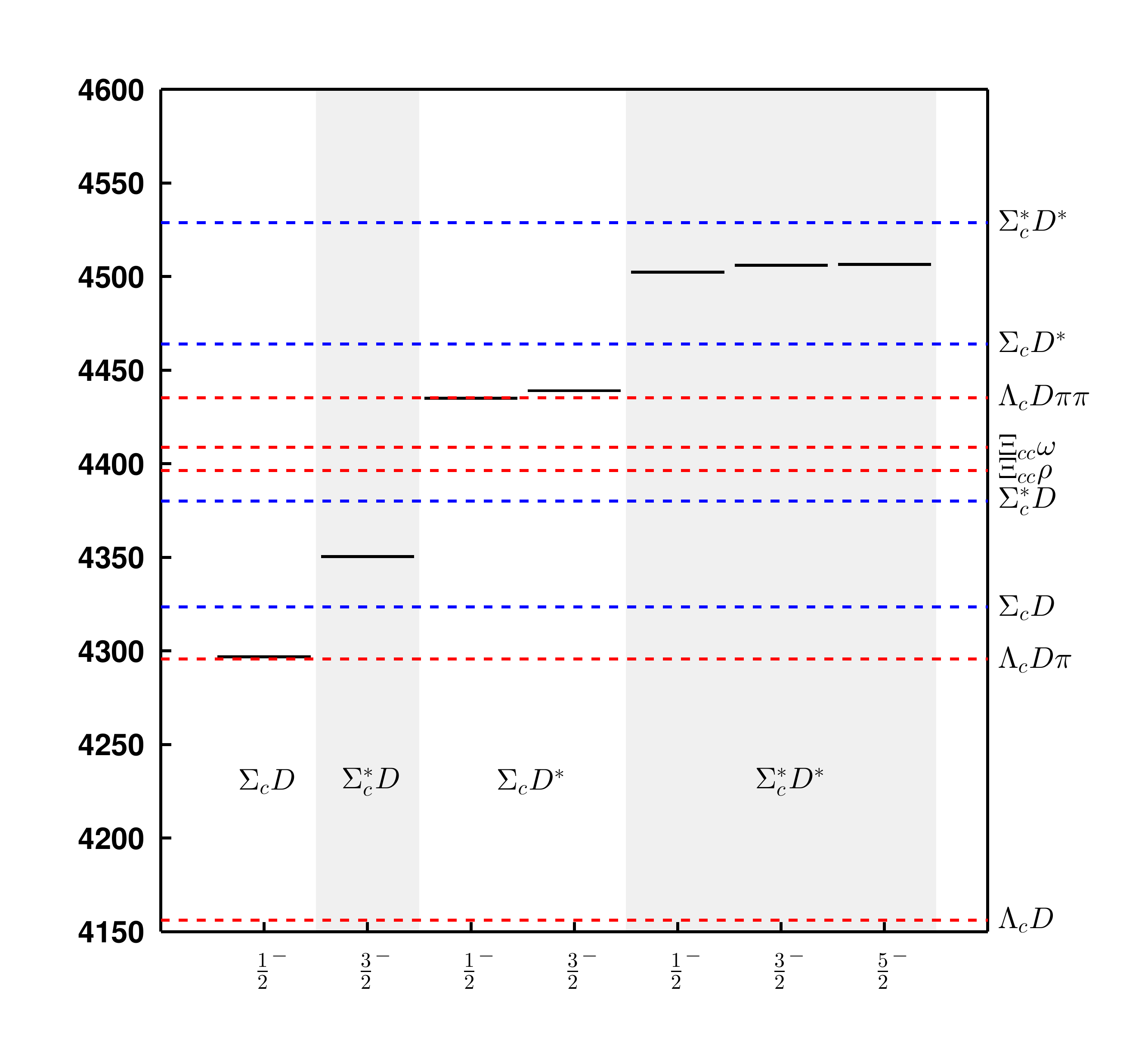}\\
\caption{The mass spectrum of the $P_{cc}$ pentaquarks. The results
are obtained using the LECs in Case 3. The black solid lines denote
the $P_{cc}$ pentaquarks. The blue and red dotted lines denote the
thresholds of the $\Sigma^{(*)}_cD^{(*)}$ and possible decay
channels, respectively. }\label{decay}
\end{figure}

In Fig. \ref{decay}, we present the mass spectrum of the $P_{cc}$
pentaquarks based on the inputs in Case 3, and some relevant
thresholds. Due to the $cc$ pairs in the $P_{cc}$ pentaquarks, the
decay behaviors of the $P_{cc}$ states are different from that of
the hidden-charm pentaquarks. There exist two types of decay modes
for the $P_{cc}$ states, i.e., the $(cqq)$-$(c\bar{q})$ and
$(ccq)$-$(q\bar{q})$ modes. Note that the $\Lambda_c\pi$ and $D\pi$
are the dominant decay channels for the $\Sigma_c^{(*)}$ baryons and
$D^*$ meson, respectively. For simplicity, we only consider the
ground $\Lambda_c$, $D$, and $\pi$ as our decay final states in the
first mode. From Fig. \ref{decay}, we can see that the
$P_{cc}(4296)$ state with $J^{P}=1/2^-$ is near the threshold of the
$\Lambda_c D\pi$. Thus, it is very difficult for this state to decay
into this three-body final state due to the small phase space. But
the $P_{cc}(4296)$ can easily decay into $\Lambda_c D$ two-body
final states. Further study on the branching ratio of this decay
process is still needed. The $P_{cc}(4350)$ with $J^P=3/2^-$ can
decay into the $\Lambda_c D\pi$ via the $S$-wave, while decaying
into the $\Lambda_c D$ is $D$-wave suppressed. One can perform
similar analyses for the other five $P_{cc}$ states.

Now we discuss the $(ccq)$-$(q\bar{q})$ decay mode, i.e., the
$P_{cc}$ states decay into the ground $\Xi_{cc}$ baryon and a
pseudoscalar or a vector meson. The threshold of $\Xi_{cc}\pi$
channel is about 3760 MeV, which is much lower than the $P_{cc}$
states and is not presented in Fig. \ref{decay}. The predicted
$P_{cc}$ states with $J^P=1/2^-$ can decay into this channel through
$S$-wave, thus, this should be an important strong decay channels
for the $J^{P}=1/2^-$ states due to the large phase spaces. The
states that are composed of the $\Sigma^{(*)}_{c}$ and $D^{*}$ can
also decay into the $\Xi_{cc}\omega$ and $\Xi_{cc}\rho$ final
states. One can also extract the decay properties for the other
$P_{cc}$ pentaquarks in Fig. \ref{decay}.

\section{Summary}\label{sec4}
Inspired by the recently observed $P_c$ \cite{Aaij:2019vzc} and
$P_{cs}$ \cite{Aaij:2020gdg} pentaquarks, we perform a systematic
study on the interactions of the $\Sigma_c^{(*)}D^{(*)}$ systems to
explore the possible $P_{cc}$ states. We include the contact term,
one-pion-exchange, and two-pion-exchange interactions within the
framework of chiral effective field theory.

Due to $G$-parity transformation law, the expressions of the
one-pion-exchange and two-pion-exchange effective potentials of the
$\Sigma_c^{(*)}D^{(*)}$ systems are opposite and identical to those
of the $\Sigma_c^{(*)}\bar{D}^{(*)}$ systems \cite{Wang:2019ato},
respectively. With the LECs fitted from the $N\bar{N}$ scattering
data, we obtain four sets of ($c_s$, $c_t$) parameters describing
the contributions of the contact terms. We present three cases to
study the binding energies of the $\Sigma_c^{(*)}D^{(*)}$ systems.

The mass spectrum of the $[\Sigma_c^{(*)}D^{(*)}]^{I=1/2}_J$
molecules depend on the values of the LECs. In Case 1, a relatively
small central potential and a large spin-spin interaction are
introduced. The obtained $P_{cc}$ mass spectrum is very similar to
that of the $\Sigma_{c}^{(*)}\bar{D}^{(*)}$ systems. However, the
mass spectra obtained in Cases 2 and 3 are different from that of
the Case 1. Further experimental studies may help us to clarify the
predicted mass spectra of the $\Sigma_{c}^{(*)}D^{(*)}$ systems in
different cases.

We briefly discuss the strong decay behaviors of the $P_{cc}$
pentaquarks. The $(cqq)$-$(c\bar{q})$ and $(ccq)$-$(q\bar{q})$ are
the two types of decay modes. Correspondingly, the $\Lambda_c D$,
$\Lambda_c D \pi$, and $\Xi_{cc}\pi$ are expected to be important
channels to search for these $[\Sigma_c^{(*)}D^{(*)}]^{I=1/2}_J$
molecules.

We also study the interactions of the $\Sigma_c^{(*)}\bar{B}^{(*)}$,
$\Sigma_{b}^{(*)}D^{(*)}$, and $\Sigma_b^{(*)}\bar{B}^{(*)}$ to
search for possible $P_{cb}$, $P_{bc}$, and $P_{bb}$ pentaquarks.
The corresponding systems with $I=1/2$ can also form molecular
states. In addition, among the studied systems, the binding becomes
deeper when the reduced masses of the systems are heavier.

\section*{Acknowledgments}
This project is supported by the National Natural Science Foundation
of China under Grants 11975033 and 12070131001.

\begin{appendix}

\section{Supplements for the two-pion-exchange expressions}\label{app1}

In Sec. \ref{Epotential}, we present the general expressions for the
football diagrams ($F_{i.j}$), triangle diagrams ($T_{i.j}$),
($\bar{T}_{i.j}$), box diagrams ($B_{i.j}$), ($\bar{B}_{i.j}$) and
crossed box diagrams $(R_{i.j})$, ($\bar{R}_{i.j}$) in Eqs.
\eqref{football}-\eqref{Lcrossedbox}. In this appendix, we give
their explicit coefficients defined in Eqs.
\eqref{football}-\eqref{Lcrossedbox}.

Specifically, we collect the coefficients
$\mathcal{C}_{\rm{sys}}^{T_{i.j}}$
($\mathcal{C}_{\rm{sys}}^{\bar{T}_{i.j}}$),
$\mathcal{C}_{\rm{sys}}^{(B/R)_{i.j}}$
($\mathcal{C}_{\rm{sys}}^{(\bar{B}/\bar{R})_{i.j}}$) in Table
\ref{Csys}, the coefficients of triangle diagrams
$\mathcal{C}_m^{T_{i.j}}$ ($\mathcal{C}_m^{\bar{T}_{i.j}}$) and
$\mathcal{E}^{T_{i.j}}$ ($\mathcal{E}^{\bar{T}_{i.j}}$) in Table
\ref{tri}, the coefficients of box and crossed box diagrams
$\mathcal{C}_m^{B_{i.j}}$, $\mathcal{C}_m^{R_{i.j}}$ and
$\mathcal{E}_n^{(B/R)_{i.j}}$ in Table \ref{tbox}, and the
coefficients of box and crossed box diagrams
$\mathcal{C}_m^{\bar{B}_{i.j}}$, $\mathcal{C}_m^{\bar{R}_{i.j}}$ and
$\mathcal{E}_n^{(\bar{B}/\bar{R})_{i.j}}$ in Table \ref{tLbox}.
\begin{table*}[!htbp]
\centering \caption{The coefficients
$\mathcal{C}_{\rm{sys}}^{T_{i.j}}$
($\mathcal{C}_{\rm{sys}}^{\bar{T}_{i.j}}$),
$\mathcal{C}_{\rm{sys}}^{(B/R)_{i.j}}$
($\mathcal{C}_{\rm{sys}}^{(\bar{B}/\bar{R})_{i.j}}$) defined in Eqs.
\eqref{triangle}-\eqref{Lcrossedbox}. The superscripts denote the
corresponding diagrams illustrated in Figs. \ref{SigmaDfig},
\ref{SigmastDfig}, \ref{SigmaDstfig}, and \ref{SigmastDstfig},
respectively. }\label{Csys}
\renewcommand\arraystretch{1.5}
\setlength{\tabcolsep}{1.3mm}{
\begin{tabular}{cccccccccccccccc}
\toprule[1pt]
&$\mathcal{C}_{\rm{sys}}^{T_{i.1}}$&$\mathcal{C}_{\rm{sys}}^{T_{i.2}}$&$\mathcal{C}_{\rm{sys}}^{T_{i.3}}$&$\mathcal{C}_{\rm{sys}}^{T_{i.4}}$&$\mathcal{C}_{\rm{sys}}^{(B/R)_{i.1}}$&$\mathcal{C}_{\rm{sys}}^{(B/R)_{i.2}}$&$\mathcal{C}_{\rm{sys}}^{(B/R)_{i.3}}$
&$\mathcal{C}_{\rm{sys}}^{(B/R)_{i.4}}$&$\mathcal{C}_{\rm{sys}}^{\bar{T}_{i.3}}$&$\mathcal{C}_{\rm{sys}}^{(\bar{B}/\bar{R})_{i.1}}$&$\mathcal{C}_{\rm{sys}}^{(\bar{B}/\bar{R})_{i.2}}$
&$\mathcal{C}_{\rm{sys}}^{(\bar{B}/\bar{R})_{i.3}}$&$\mathcal{C}_{\rm{sys}}^{(\bar{B}/\bar{R})_{i.4}}$\\
\hline
$\Sigma_c D$&$g^2$&$\frac{g_3^2}{4}$&$\frac{g_1^2}{4}$&-&$\frac{g^2g_1^2}{8}$&$\frac{g^2g_3^2}{8}$&-&-&$\frac{g_2^2}{2}$&$\frac{g^2g_2^2}{8}$&-&-&-\\
$\Sigma_c^* D$&$g^2$&$\frac{5g_5^2}{36}$&$\frac{g_3^2}{4}$&-&$\frac{5g^2g_5^2}{72}$&$\frac{g^2g_3^2}{8}$&-&-&$\frac{g_4^2}{2}$&-&$\frac{g^2g_4^2}{8}$&-&-\\
$\Sigma_c D^*$&$g^2$&$g^2$&$\frac{g_1^2}{4}$&$\frac{g_3^2}{4}$&$\frac{g^2g_1^2}{8}$&$\frac{g^2g_1^2}{8}$&$\frac{g^2g_3^2}{8}$&$\frac{g^2g_3^2}{24}$&$\frac{g_2^2}{2}$&$\frac{g^2g_2^2}{8}$&$\frac{g^2g_2^2}{8}$&-&-\\
$\Sigma_c^* D^*$&$g^2$&$g^2$&$\frac{g_3^2}{4}$&$\frac{5g_5^2}{36}$&$\frac{g^2g_5^2}{24}$&$\frac{g^2g_5^2}{24}$&$\frac{g^2g_3^2}{32}$&$\frac{g^2g_3^2}{32}$&$\frac{g_4^2}{2}$&-&-& $\frac{g^2g_4^2}{32}$&$\frac{g^2g_4^2}{32}$\\
\bottomrule[1pt]
\end{tabular}
}
\end{table*}


\begin{table*}[!htbp]
\centering \caption{The coefficients $\mathcal{C}_m^{T_{i.j}}$
($\mathcal{C}_m^{\bar{T}_{i.j}}$) and $\mathcal{E}^{T_{i.j}}$
($\mathcal{E}^{\bar{T}_{i.j}}$) defined in Eqs. \eqref{triangle} and
\eqref{Ltriangle}. The superscript denotes the corresponding
diagrams illustrated in Figs. \ref{SigmaDfig}, \ref{SigmastDfig},
\ref{SigmaDstfig}, and \ref{SigmastDstfig}, respectively.
}\label{tri}
\renewcommand\arraystretch{1.4}
\setlength{\tabcolsep}{1.85mm}{
\begin{tabular}{cccccccccccccccccccccccccccccc}
\toprule[1pt]
&$\mathcal{C}_1^{T_{i.1}}$&$\mathcal{C}_2^{T_{i.1}}$&$\mathcal{E}^{T_{i.1}}$
&$\mathcal{C}_1^{T_{i.2}}$&$\mathcal{C}_2^{T_{i.2}}$&$\mathcal{E}^{T_{i.2}}$&$\mathcal{C}_1^{T_{i.3}}$&$\mathcal{C}_2^{T_{i.3}}$&$\mathcal{E}^{T_{i.3}}$
&$\mathcal{C}_1^{T_{i.4}}$&$\mathcal{C}_2^{T_{i.4}}$&$\mathcal{E}^{T_{i.4}}$&$\mathcal{C}_1^{\bar{T}_{i.3}}$&$\mathcal{C}_2^{\bar{T}_{i.3}}$&$\mathcal{E}^{\bar{T}_{i.3}}$\\
\hline
$\Sigma_c D$&1&3&$\mathcal{E}-\delta_b$&$\frac{2}{3}$&2&$\mathcal{E}-\delta_a$&1&3&$\mathcal{E}$&-&-&-&1&3&$\mathcal{E}+\delta_c$\\
$\Sigma_c^* D$&1&3&$\mathcal{E}-\delta_b$&1&3&$\mathcal{E}$&$\frac{1}{3}$&1&$\mathcal{E}+\delta_a$&-&-&-&$\frac{1}{3}$&1&$\mathcal{E}+\delta_d$\\
$\Sigma_cD^*$&$\frac{2}{3}$&2&$\mathcal{E}$&$\frac{1}{3}$&1&$\mathcal{E}+\delta_b$&1&3&$\mathcal{E}$&$\frac{2}{3}$&2&$\mathcal{E}-\delta_a$&1&3&$\mathcal{E}+\delta_c$\\
$\Sigma_c^* D^*$&$\frac{2}{3}$&2&$\mathcal{E}$&$\frac{1}{3}$&1&$\mathcal{E}+\delta_b$&$\frac{1}{3}$&1&$\mathcal{E}+\delta_a$&1&3&$\mathcal{E}$&$\frac{1}{3}$&1&$\mathcal{E}+\delta_d$\\
\bottomrule[1pt]
\end{tabular}
}
\end{table*}

\begin{table*}[!htbp]
\centering \caption{The coefficients $\mathcal{C}_m^{B_{i.j}}$,
$\mathcal{C}_m^{R_{i.j}}$ and $\mathcal{E}_n^{(B/R)_{i.j}}$ defined
in Eqs. \eqref{box} and \eqref{crossedbox}. The superscripts denote
the corresponding diagrams illustrated in Figs. \ref{SigmaDfig},
\ref{SigmastDfig}, \ref{SigmaDstfig}, and \ref{SigmastDstfig},
respectively. }\label{tbox}
\renewcommand\arraystretch{1.3}
\setlength{\tabcolsep}{3.9mm}{
\begin{tabular}{cccccccccccccccccccccccccccccc}
\toprule[1pt]
&$\mathcal{C}_1^{B_{i.1}}$&$\mathcal{C}_1^{R_{i.1}}$&$\mathcal{C}_2^{(B/R)_{i.1}}$&$\mathcal{C}_{3}^{(B/R)_{i.1}}$&$\mathcal{C}_{4}^{(B/R)_{i.1}}$&$\mathcal{E}_1^{(B/R)_{i.1}}$&$\mathcal{E}_2^{(B/R)_{i.1}}$\\
\hline
$\Sigma_cD$&1&1&1&10&15&$\mathcal{E}$&$\mathcal{E}-\delta_b$\\
$\Sigma_c^*D$&1&1&1&10&15&$\mathcal{E}$&$\mathcal{E}-\delta_b$\\
$\Sigma_cD^*$&$\frac{2+A}{3}$&$\frac{2-A}{3}$&$\frac{2}{3}$&$\frac{20}{3}$&10&$\mathcal{E}$&$\mathcal{E}$\\
$\Sigma^*_cD^*$&$B^2$&$\frac{3B^2-2B}{3}$&$\frac{10}{9}$
&$\frac{20+12B^2-4B}{3}$&$10+6B^2-2B$&$\mathcal{E}$&$\mathcal{E}$\\
\hline
&$\mathcal{C}_1^{B_{i.2}}$&$\mathcal{C}_1^{R_{i.2}}$&$\mathcal{C}_2^{(B/R)_{i.2}}$&$\mathcal{C}_{3}^{(B/R)_{i.2}}$&$\mathcal{C}_{4}^{(B/R)_{i.2}}$&$\mathcal{E}_1^{(B/R)_{i.2}}$&$\mathcal{E}_1^{(B/R)_{i.2}}$\\
\hline
$\Sigma_cD$&$\frac{2}{3}$&$\frac{2}{3}$&$\frac{2}{3}$&$\frac{20}{3}$&10&$\mathcal{E}-\delta_a$&$\mathcal{E}-\delta_b$\\
$\Sigma_c^*D$&$\frac{1}{3}$&$\frac{1}{3}$&$\frac{1}{3}$&$\frac{10}{3}$&5&$\mathcal{E}+\delta_a$&$\mathcal{E}-\delta_b$\\
$\Sigma_cD^*$&$\frac{1+A}{3}$&$\frac{1-A}{3}$&$\frac{1}{3}$&$\frac{10}{3}$&5&$\mathcal{E}$&$\mathcal{E}+\delta_b$\\
$\Sigma^*_cD^*$&$\frac{5-3B^2+2B}{3}$&$\frac{5-3B^2}{3}$
&$\frac{5}{9}$&$\frac{15-6B^2+2B}{3}$&$15-6B^2+2B$&$\mathcal{E}$&$\mathcal{E}+\delta_b$\\
\hline
&$\mathcal{C}_1^{B_{i.3}}$&$\mathcal{C}_1^{R_{i.3}}$&$\mathcal{C}_2^{(B/R)_{i.3}}$&$\mathcal{C}_{3}^{(B/R)_{i.3}}$&$\mathcal{C}_{4}^{(B/R)_{i.3}}$&$\mathcal{E}_1^{(B/R)_{i.3}}$&$\mathcal{E}_1^{(B/R)_{i.3}}$\\
\hline
$\Sigma_cD^*$&$\frac{4-A}{9}$&$\frac{4+A}{9}$&$\frac{4}{9}$&$\frac{40}{9}$&$\frac{20}{3}$&$\mathcal{E}-\delta_a$&$\mathcal{E}$\\
$\Sigma^*_cD^*$&$2-B^2+B$&$2-B^2-\frac{B}{3}$&$\frac{8}{9}$
&$\frac{40-12B^2+4B}{3}$&$20-6B^2+2B$&$\mathcal{E}+\delta_a$&$\mathcal{E}$\\
\hline
&$\mathcal{C}_1^{B_{i.4}}$&$\mathcal{C}_1^{R_{i.4}}$&$\mathcal{C}_2^{(B/R)_{i.4}}$&$\mathcal{C}_{3}^{(B/R)_{i.4}}$&$\mathcal{C}_{4}^{(B/R)_{i.4}}$&$\mathcal{E}_1^{(B/R)_{i.4}}$&$\mathcal{E}_1^{(B/R)_{i.4}}$\\
\hline
$\Sigma_cD^*$&$\frac{2-A}{3}$&$\frac{2+A}{3}$&$\frac{2}{3}$&$\frac{20}{3}$&10&$\mathcal{E}-\delta_a$&$\mathcal{E}+\delta_b$\\
$\Sigma^*_cD^*$&$\frac{3B^2+B-2}{3}$&$\frac{3B^2-3B-2}{3}$&$\frac{4}{9}$
&$\frac{12B^2-4B}{3}$&$6B^2-2B$&$\mathcal{E}+\delta_a$&$\mathcal{E}+\delta_b$\\
\bottomrule[1pt]
\end{tabular}
}
\end{table*}

\begin{table*}[t]
\centering \caption{The coefficients
$\mathcal{C}_m^{\bar{B}_{i.j}}$, $\mathcal{C}_m^{\bar{R}_{i.j}}$ and
$\mathcal{E}_n^{(\bar{B}/\bar{R})_{i.j}}$ defined in Eqs.
\eqref{Lbox} and \eqref{Lcrossedbox}. The superscripts denote the
corresponding diagrams illustrated in Figs. \ref{SigmaDfig},
\ref{SigmastDfig}, \ref{SigmaDstfig}, and \ref{SigmastDstfig},
respectively. }\label{tLbox}
\renewcommand\arraystretch{1.5}
\setlength{\tabcolsep}{3.85mm}{
\begin{tabular}{cccccccccccccccccccccccccccccc}
\toprule[1pt]
&$\mathcal{C}_1^{\bar{B}_{1.1}}$&$\mathcal{C}_1^{\bar{R}_{1.1}}$&$\mathcal{C}_2^{(\bar{B}/\bar{R})_{1.1}}$&$\mathcal{C}_{3}^{(\bar{B}/\bar{R})_{1.1}}$&$\mathcal{C}_{4}^{(\bar{B}/\bar{R})_{1.1}}$&$\mathcal{E}_1^{(\bar{B}/\bar{R})_{1.1}}$&$\mathcal{E}_2^{(\bar{B}/\bar{R})_{1.1}}$\\
\hline
$\Sigma_cD$&1&1&1&10&15&$\mathcal{E}+\delta_c$&$\mathcal{E}-\delta_b$\\
\hline
&$\mathcal{C}_1^{\bar{B}_{2.2}}$&$\mathcal{C}_1^{\bar{R}_{2.2}}$&$\mathcal{C}_2^{(\bar{B}/\bar{R})_{2.2}}$&$\mathcal{C}_{3}^{(\bar{B}/\bar{R})_{2.2}}$&$\mathcal{C}_{4}^{(\bar{B}/\bar{R})_{2.2}}$&$\mathcal{E}_1^{(\bar{B}/\bar{R})_{2.2}}$&$\mathcal{E}_2^{(\bar{B}/\bar{R})_{2.2}}$\\
\hline
$\Sigma_c^*D$&$\frac{1}{3}$&$\frac{1}{3}$&$\frac{1}{3}$&$\frac{10}{3}$&5&$\mathcal{E}+\delta_d$&$\mathcal{E}-\delta_b$\\
\hline
&$\mathcal{C}_1^{\bar{B}_{3.1}}$&$\mathcal{C}_1^{\bar{R}_{3.1}}$&$\mathcal{C}_2^{(\bar{B}/\bar{R})_{3.1}}$&$\mathcal{C}_{3}^{(\bar{B}/\bar{R})_{3.1}}$&$\mathcal{C}_{4}^{(\bar{B}/\bar{R})_{3.1}}$&$\mathcal{E}_1^{(\bar{B}/\bar{R})_{3.1}}$&$\mathcal{E}_2^{(\bar{B}/\bar{R})_{3.1}}$\\
\hline
$\Sigma_cD^*$&$\frac{2+A}{3}$&$\frac{2-A}{3}$&$\frac{2}{3}$&$\frac{20}{3}$&10&$\mathcal{E}+\delta_c$&$\mathcal{E}$\\
\hline
&$\mathcal{C}_1^{\bar{B}_{3.2}}$&$\mathcal{C}_1^{\bar{R}_{3.2}}$&$\mathcal{C}_2^{(\bar{B}/\bar{R})_{3.2}}$&$\mathcal{C}_{3}^{(\bar{B}/\bar{R})_{3.2}}$&$\mathcal{C}_{4}^{(\bar{B}/\bar{R})_{3.2}}$&$\mathcal{E}_1^{(\bar{B}/\bar{R})_{3.2}}$&$\mathcal{E}_2^{(\bar{B}/\bar{R})_{3.2}}$\\
\hline
$\Sigma_cD^*$&$\frac{1+A}{3}$&$\frac{1-A}{3}$&$\frac{1}{3}$&$\frac{10}{3}$&5&$\mathcal{E}+\delta_c$&$\mathcal{E}+\delta_b$\\
\hline
&$\mathcal{C}_1^{\bar{B}_{4.3}}$&$\mathcal{C}_1^{\bar{R}_{4.3}}$&$\mathcal{C}_2^{(\bar{B}/\bar{R})_{4.3}}$&$\mathcal{C}_{3}^{(\bar{B}/\bar{R})_{4.3}}$&$\mathcal{C}_{4}^{(\bar{B}/\bar{R})_{4.3}}$&$\mathcal{E}_1^{(\bar{B}/\bar{R})_{4.3}}$&$\mathcal{E}_2^{(\bar{B}/\bar{R})_{4.3}}$\\
\hline
$\Sigma^*_cD^*$&$2-B^2+B$&$2-B^2-\frac{B}{3}$&$\frac{8}{9}$&$\frac{40-12B^2+4B}{3}$&$20-6B^2+2B$
&$\mathcal{E}+\delta_d$&$\mathcal{E}$\\
\hline
&$\mathcal{C}_1^{\bar{B}_{4.4}}$&$\mathcal{C}_1^{\bar{R}_{4.4}}$&$\mathcal{C}_2^{(\bar{B}/\bar{R})_{4.4}}$&$\mathcal{C}_{3}^{(\bar{B}/\bar{R})_{4.4}}$&$\mathcal{C}_{4}^{(\bar{B}/\bar{R})_{4.4}}$&$\mathcal{E}_1^{(\bar{B}/\bar{R})_{4.4}}$&$\mathcal{E}_2^{(\bar{B}/\bar{R})_{4.4}}$\\
\hline
$\Sigma_c^*D^*$&$\frac{3B^2+B-2}{3}$&$\frac{3B^2-3B-2}{3}$&$\frac{4}{9}$&
$\frac{3B^2-B}{3}$&$6B^2-2B$&$\mathcal{E}+\delta_d$&$\mathcal{E}+\delta_b$\\
\bottomrule[1pt]
\end{tabular}
}
\end{table*}

\section{The binding energies of the $\Sigma_c^{(*)}\bar{B}^{(*)}$, $\Sigma_b^{(*)}D^{(*)}$, and $\Sigma_b^{(*)}\bar{B}^{(*)}$ systems}\label{app2}

We also study the interactions of the $\Sigma_c^{(*)}\bar{B}^{(*)}$,
$\Sigma_b^{(*)}D^{(*)}$, and $\Sigma_b^{(*)}\bar{B}^{(*)}$ systems
with the three cases of LECs. The results for the possible $P_{cb}$
($\Sigma_c^{(*)}\bar{B}^{(*)}$), $P_{bc}$ ($\Sigma_b^{(*)}D^{(*)}$),
and $P_{bb}$ ($\Sigma_b^{(*)}\bar{B}^{(*)}$) pentaquarks are
collected in Tables \ref{Pcb}, \ref{Pbc}, and \ref{Pbb},
respectively.

In our calculation, we have already adopted the approach developed
in Ref. \cite{Wang:2019ato} to keep the effects from the mass
splittings in the ($\Sigma_b^*$, $\Sigma_b$, $\Lambda_b$) baryons
and ($\bar{B}^*$, $B$) mesons.

We find binding solutions for all the
$[\Sigma_c^{(*)}\bar{B}^{(*)}]^{I=1/2}_J$,
$[\Sigma_b^{(*)}D^{(*)}]^{I=1/2}_J$, and
$[\Sigma_b^{(*)}\bar{B}^{(*)}]^{I=1/2}_J$ systems. The mass spectra
of the possible $P_{cb}$, $P_{bc}$, and $P_{bb}$ pentaquarks are
similar to those of the $\Sigma_c^{(*)}D^{(*)}$ systems.

We also notice that among the $[\Sigma_c^{(*)}D^{(*)}]^{I=1/2}_J$,
$[\Sigma_c^{(*)}\bar{B}^{(*)}]^{I=1/2}_J$,
$[\Sigma_b^{(*)}D^{(*)}]^{I=1/2}_J$, and
$[\Sigma_b^{(*)}\bar{B}^{(*)}]^{I=1/2}_J$ systems, the absolute
values of the binding energies generally have the following
relation,
\begin{eqnarray}
|E_{P_{cc}}|<|E_{P_{cb}}|\approx|E_{P_{bc}}|<|E_{P_{bb}}|.
\end{eqnarray}

\begin{table*}[!htbp]
\caption{The binding energies and root-mean-equare radii for all the
$[\Sigma_c^{(*)}\bar{B}^{(*)}]^{I=1/2}_J$ systems. The adopted LECs
in Cases 1, 2, and 3 are ($c_s=-5.84$, $c_t=2.50$) GeV$^{-2}$,
($c_s=-8.10$, $c_t=0.65$ ) GeV$^{-2}$, and ($c_s=-7.46$, $c_t=1.02$)
GeV$^{-2}$, respectively.}\label{Pcb}
\renewcommand\arraystretch{1.15}
\setlength{\tabcolsep}{3.85mm}{
\begin{tabular}{c|c|ccccccccccccccccc}
\toprule[1pt] &&$\left[\Sigma_c\bar{B}\right]_{\frac{1}{2}}$
&$\left[\Sigma^*_c\bar{B}\right]_{\frac{3}{2}}$&$\left[\Sigma_c\bar{B}^*\right]_{\frac{1}{2}}$&$\left[\Sigma_c\bar{B}^*\right]_{\frac{3}{2}}$& $\left[\Sigma_c^*\bar{B}^*\right]_{\frac{1}{2}}$&$\left[\Sigma_c^*\bar{B}^*\right]_{\frac{3}{2}}$&$\left[\Sigma_c^*\bar{B}^*\right]_{\frac{5}{2}}$\\
\hline
\multirow{2}{*}{Case 1} &BE (MeV)&-24.3&-24.8&-39.8&-13.0&-47.6&-32.7&-11.0\\
&$R_{rms}$ (fm)&1.16&1.15&1.04&1.35&1.00&1.08&1.38\\
\hline
\multirow{2}{*}{Case 2} &BE (MeV)&-43.5&-44.0&-38.0&-40.9&-40.3&-41.5&-44.1\\
&$R_{rms}$ (fm)&1.01&1.01&1.06&1.03&1.04&1.03&1.01\\
\hline
\multirow{2}{*}{Case 3} &BE (MeV)&-37.9&-44.9&-36.8&-33.3&-40.1&-38.1&-35.3\\
&$R_{rms}$ (fm)&1.05&1.00&1.06&1.08&1.04&1.05&1.06\\
\bottomrule[1pt]
\end{tabular}
}
\end{table*}

\begin{table*}[!htbp]
\caption{The binding energies and root-mean-equare radii for all the
$[\Sigma_b^{(*)}D^{(*)}]_J^{I=1/2}$ systems. The adopted LECs in
Cases 1, 2, and 3 are ($c_s=-5.84$, $c_t=2.50$) GeV$^{-2}$,
($c_s=-8.10$, $c_t=0.65$ ) GeV$^{-2}$, and ($c_s=-7.46$, $c_t=1.02$)
GeV$^{-2}$, respectively.}\label{Pbc}
\renewcommand\arraystretch{1.15}
\setlength{\tabcolsep}{4.12mm}{
\begin{tabular}{c|c|ccccccccccccccccc}
\toprule[1pt] &&$\left[\Sigma_bD\right]_{\frac{1}{2}}$
&$\left[\Sigma^*_bD\right]_{\frac{3}{2}}$&$\left[\Sigma_bD^*\right]_{\frac{1}{2}}$&$\left[\Sigma_bD^*\right]_{\frac{3}{2}}$& $\left[\Sigma_b^*D^*\right]_{\frac{1}{2}}$&$\left[\Sigma_b^*D^*\right]_{\frac{3}{2}}$&$\left[\Sigma_b^*D^*\right]_{\frac{5}{2}}$\\
\hline
\multirow{2}{*}{Case 1} &BE (MeV)&-23.2&-21.2&-37.3&-11.5&-43.0&-28.0&-7.8\\
&$R_{rms}$ (fm)&1.21&1.25&1.09&1.42&1.05&1.16&1.55\\
\hline
\multirow{2}{*}{Case 2} &BE (MeV)&-41.7&-39.3&-35.6&-38.4&-36.0&-36.5&-38.6\\
&$R_{rms}$ (fm)&1.06&1.08&1.10&1.07&1.10&1.10&1.07\\
\hline
\multirow{2}{*}{Case 3} &BE (MeV)&-36.3&-34.0&-34.4&-31.0&-35.8&-33.2&-30.2\\
&$R_{rms}$ (fm)&1.10&1.12&1.11&1.13&1.10&1.12&1.13\\
\bottomrule[1pt]
\end{tabular}
}
\end{table*}

\begin{table*}[!htbp]
\caption{The binding energies and root-mean-equare radii for all the
$[\Sigma_b^{(*)}\bar{B}^{(*)}]_J^{I=1/2}$ systems. The adopted LECs
in Cases 1, 2, and 3 are ($c_s=-5.84$, $c_t=2.50$) GeV$^{-2}$,
($c_s=-8.10$, $c_t=0.65$ ) GeV$^{-2}$, and ($c_s=-7.46$, $c_t=1.02$)
GeV$^{-2}$, respectively.}\label{Pbb}
\renewcommand\arraystretch{1.15}
\setlength{\tabcolsep}{3.85mm}{
\begin{tabular}{c|c|ccccccccccccccccc}
\toprule[1pt] &&$\left[\Sigma_b\bar{B}\right]_{\frac{1}{2}}$
&$\left[\Sigma^*_b\bar{B}\right]_{\frac{3}{2}}$&$\left[\Sigma_b\bar{B}^*\right]_{\frac{1}{2}}$&$\left[\Sigma_b\bar{B}^*\right]_{\frac{3}{2}}$& $\left[\Sigma_b^*\bar{B}^*\right]_{\frac{1}{2}}$&$\left[\Sigma_b^*\bar{B}^*\right]_{\frac{3}{2}}$&$\left[\Sigma_b^*\bar{B}^*\right]_{\frac{5}{2}}$\\
\hline
\multirow{2}{*}{Case 1} &BE (MeV)&-33.2&-30.5&-53.0&-20.0&-60.0&-41.8&-15.1\\
&$R_{rms}$ (fm)&0.97&0.99&0.87&1.09&0.85&0.92&1.16\\
\hline
\multirow{2}{*}{Case 2} &BE (MeV)&-54.8&-53.5&-51.0&-52.2&-51.5&-51.7&-52.5\\
&$R_{rms}$ (fm)&0.86&0.87&0.88&0.87&0.88&0.87&0.87\\
\hline
\multirow{2}{*}{Case 3} &BE (MeV)&-48.5&-46.7&-49.5&-43.6&-51.3&-47.9&-42.7\\
&$R_{rms}$ (fm)&0.88&0.89&0.89&0.91&0.88&0.89&0.91\\
\bottomrule[1pt]
\end{tabular}
}
\end{table*}


\end{appendix}

\end{document}